\documentclass[final, 12 pt]{elsarticle}

\usepackage{lineno,hyperref}

\journal{peer-reviewed journal.}

\usepackage{tikz} 
\usetikzlibrary{matrix}

\usepackage{pgfplots, pgfplotstable}
\usepgfplotslibrary{statistics}
\usepgfplotslibrary{groupplots}
\pgfplotsset{compat=newest}
\pgfplotsset{every axis/.append style={
                    label style={font=\scriptsize},
                    tick label style={font=\scriptsize},
                    legend style={font=\scriptsize},
                    }}

\usepackage{subcaption}
\usepackage{siunitx}
\usepackage{booktabs}
\usepackage{multirow}
\usepackage{amsmath}
\usepackage{amssymb}
\usepackage{paralist}

\begin{document}

\begin{frontmatter}

\title{Correlation structure in the elasticity tensor for short fiber-reinforced composites}

\author[mysecondaryaddress]{Natalie Rauter\corref{mycorrespondingauthor}}
\cortext[mycorrespondingauthor]{Corresponding author}
\ead{natalie.rauter@hsu-hh.de}

\author[mysecondaryaddress]{Rolf Lammering}

\address[mysecondaryaddress]{Helmut-Schmidt-University / University of the Federal Armed Forces Hamburg,\\ 22043 Hamburg, Germany}

\begin{abstract}
The present work provides a profound analytical and numerical analysis of the material properties of SFRC on the mesoscale as well as the resulting correlation structure taking into account the probabilistic characteristics of the fiber geometry. This is done by calculating the engineering constants using the analytical framework given by Tandon and Weng as well as Halpin and Tsai. The input parameters like fiber length, diameter and orientation are chosen with respect to their probability density function. It is shown, that they are significantly influenced by the fiber length, the fiber orientation and the fiber volume fraction. The verification of the analytically obtained values is done on a numerical basis. Therefore, a two-dimensional microstructure is generated and transferred to a numerical model. The advantage of this procedure is, that there are several fibers with different geometrical properties placed in a preset area. The results of the numerical analysis meet the analytically obtained conclusions. Furthermore, the results of the numerical simulations are independent of the assumption of a plane strain and plane stress state, respectively. Finally, the correlation structure of the elasticity tensor is investigated. Not only the symmetry properties of the elasticity tensor characterize the correlation structure, but also the overall transversely-isotropic material behavior is confirmed. In contrast to the influencing parameters, the correlation functions vary for a plane strain and a plane stress state.
\end{abstract}

\begin{keyword}
Short fiber-reinforced composites, Correlation analysis, Moving window
\end{keyword}

\end{frontmatter}


\section{Introduction}
Short fiber-reinforced composites (SFRC) are widely used in the automotive and aeronautical industry. One main advantage is the suitability of thermoplastic-based compounds for an automated serial production like mold injection, which allows high production rates with reasonable prices per piece. However, due to the finite fiber length and varying flow velocities and directions during the injection processes, the components show spatially distributed mechanical properties. Due to this the corresponding numerical simulation of the components is challenging and expensive. For an adequate material description, representative material properties need to be established. 

From an analytic perspective in contrast to continuous fiber-reinforced composites, the local effects on stress and strain states due to microscopic inclusions must be taken into account. One approach for an analytic description of the resulting mechanical properties of SFRC is based on the mean-field theory in combination with Eshelby's work \cite{Eshelby.1957}. Mori and Tanaka \cite{Mori.1973} and later Tandon and Weng \cite{Tandon.1984} expanded this. As a second approach self-consistent models are introduced. One very common representation of this group was developed by Halpin and Tsai \cite{Halpin.1969,Halpin.1976}. A detailed overview of the analytic modeling is given in \cite{TuckerIII.1999}. It is concluded that the approach by Tandon and Weng shows the best results for the prediction of the elastic properties. This is also confirmed in \cite{Gusev.2000}. However, as these material models are based on homogeneous material they are not capable of representing the spatial dependence of the material properties.

There are various analyses of the influencing parameters based on analytical material models. In \cite{Gusev.2000} it is shown that for example the spatial fiber distribution must be taken into account. The effect of the fiber orientation on an analytical basis is given in \cite{Fu.1998}. Finally, in \cite{Huang.2001} a comprehensive analysis including not only the fiber orientation and fiber distribution but also the aspect ratio of the fiber is taken into account. However, these studies don't include the probability properties of the fiber characteristics.

Following the analytical analyses of the influencing parameters, the resulting fluctuation of the fiber volume fraction and fiber orientation should be added to numerical models of SFRC. To represent the spatial fiber distribution the fiber itself must be added to the model. However, as the fiber length is very small compared to component dimensions a multi-scale approach is required for numerical modeling. One possibility here is the use of second-order random fields, that represent spatial varying information on the mesoscale \cite{Guilleminot.2008} and therefore, allow the modeling of inhomogeneous material properties \cite{OstojaStarzewski.1999}. For example in \cite{Zimmermann.2018} this technique is used to simulate numerically the continuous mode conversion of Lamb waves in fiber reinforced composite structures, which is induced by the random distribution of the fibers on the microscale.

Other approaches for the numerical modeling of SFRC are based on the representative volume element (RVE). For example, in \cite{Hickmann.2016} the suitability of randomly generated characteristic volume elements using XFEM is analyzed to capture the local material response in concrete. XFEM is also used in \cite{Pike.2015} in combination with a cohesive zone model. In \cite{Zairi.2008} micro-mechanical modeling of randomly oriented fiber polymer composites is presented. Furthermore, an overview of studies deriving a micromechanical model for randomly orientated fibers is given. The suitability of the orientation tensor is discussed in \cite{Behrens.2014,Muller.2016}.

The main goal of this research is to evaluate profoundly the microstructural properties of SFRC with respect to the probabilistic characteristic on both an analytical as well as on a numerical basis. These properties can later be used to generate second-order random fields for an adequate representation of the material properties on the component level. In a first step the analytic models of SFRC are used to analyze the influence of the main material characteristics like fiber length, fiber orientation, fiber diameter, and fiber volume fraction on the elasticity properties. Therefore, each characteristic is separately varied following its known probability density functions (PDF) and the elasticity parameters are calculated. Afterward, the results of all analytically obtained material properties are compared with each other to identify the characteristics influencing the material properties most. This analytic analysis is extended to two-dimensional numerical simulations, which is based on a predefined area containing randomly placed reinforcing fibers. The analytically obtained results are used to verify the numerical model by comparing the calculated engineering constants and elasticity coefficients.

For investigating the influence of the main characteristics again each characteristic is analyzed separately. Besides this, an additional simulation is carried out where all parameters are considered as varying. The simulations are performed under boundary conditions of Neumann and Dirichlet type, which allows to obtain an upper (Voigt) and a lower (Reuss) bound. From these analyses, the influence of the main material characteristics on the elasticity properties for SFRC are derived. Based on these results further numerical simulations are carried out. Now the main focus lies on the spatial dependence of the material properties. This behavior can be expressed by correlation functions. Therefore, the moving window method is used to obtain the correlation nature of the elasticity properties. A first analysis of the correlation analysis of the elasticity parameters is done in \cite{Sena.2013}. However, that analysis is limited to a checkerboard pattern and therefore, does not include effects based on the varying geometrical properties of the reinforcing fibers as well as their orientation. Therefore, in this study this approach is extended to randomly placed fibers in a predefined area.

The structure of the paper is as follows. Section 2 gives an overview of the theoretical background for multi-scale modeling of fiber-reinforced composites.  This includes the determination of effective material properties as well as the framework of correlated second-order random fields. This is followed by a profound analysis of the microstructure in Section 3, which includes first an overview of probabilistic properties of the main geometrical characteristics of SFRC followed by the description of the microstructure generation and the corresponding numerical model. Furthermore, the influence of the geometrical characteristics and their probabilistic nature on the mechanical properties is presented. In Section 4 the verification of the analytical results by numerical simulations is presented. Moreover, the numerical simulation is used to investigate the correlation structure of the elasticity tensor. Finally, Section 5 gives a summary and conclusion of the presented work.

\section{Theoretical Framework}
In this section a brief overview of the main theoretical framework used in this study is given, which covers the multi-scale modeling based on the principle of separation of scales. Regarding this approach definitions of boundary conditions are presented, that are used to determine the effective material properties of SFRC on the mesoscale. Next is a short summary of the analytical modeling of SFRC and the resulting elasticity coefficients. Finally, this section is concluded by a short introduction of second-order random fields that can be used to represent the spatial fluctuation of fiber properties on the mesoscale.

\subsection{Numerical multi-scale modeling}
\label{sec:Theory}

\subsubsection{RVE}
SFRC consists of two different components, namely the matrix material and randomly distributed embedded fibers. On this microscopic scale, reinforced materials are therefore heterogeneous. This contradicts the traditional continuum mechanics approach, which is based on homogeneous material properties independent of the volume size. Therefore, suitable techniques for the representation of microstructural inhomogeneities are necessary. One technique here is the homogenization, which takes heterogeneous properties from the microstructure into account. The goal is to define a RVE, for which the heterogeneous material properties can be replaced by homogeneous effective material properties. For the edge length $d$ of a RVE
\begin{equation}
    l \le d \le L
\end{equation}
holds. Here, $l$ is the size of an inclusion and therefore, is assigned to the microscale, whereas $L$ represents the dimension of the macroscale \cite{Hashin.1983}. This approach is also known as separation of scales. It states that on the macroscale the size $d$ can be seen as a material point, however with respect to the size of the inclusion $d$ must contain statistically representative information about the microstructure \cite{Zohdi.2010}.

A well known definition of RVE can be found in \cite{Hill.1963}. Accordingly, the dimensions of a RVE must be selected in such a way that the resulting effective material properties are independent of the boundary conditions. If the used scale is smaller than the corresponding RVE, the boundary conditions as well as the microstructure and the contrast of the different phases must be taken into account \cite{Hazanov.1994,Huet.1990}. In this case one speaks of a statistical volume element (SVE) \cite{OstojaStarzewski.2011}. The corresponding material properties are called apparent overall properties \cite{Huet.1982,Huet.1984}.

\subsubsection{Effective material properties}
The homogeneous effective material properties of a RVE can be written as 
\begin{equation}
    \langle \boldsymbol{\sigma}\rangle = \mathbb{C}^{\text{eff}} : \langle \boldsymbol{\epsilon} \rangle
    \label{eqn:Ceff}
\end{equation}
and
\begin{equation}
    \langle \boldsymbol{\epsilon}\rangle = \mathbb{S}^{\text{eff}} : \langle \boldsymbol{\sigma} \rangle
    \label{eqn:Seff}
\end{equation}
respectively. Here, the strains $\boldsymbol{\epsilon}$ and stress $\boldsymbol{\sigma}$ provide information about the microscale, whereas $\langle \cdot \rangle$ indicates the volume average, which is defined as
\begin{equation}
    \langle \cdot \rangle = \frac{1}{V} \int_V \cdot \operatorname{d}V,
\end{equation}
and therefore, represent the whole RVE. This integral relation leads to the fact, that Eqs. (\ref{eqn:Ceff}) and (\ref{eqn:Seff}) can not be inverted. Furthermore, $\mathbb{C}^{\text{eff}}$ describes the effective elasticity tensor, whereas $\mathbb{S}^{\text{eff}}$ is the effective compliance tensor.

For the determination of these effective material properties the Hill's condition is essential \cite{Hill.1952}. It states, that the conservation of energy for the scale transition can be written as
\begin{equation}
    \langle \boldsymbol{\sigma} : \boldsymbol{\epsilon} \rangle = \langle \boldsymbol{\sigma}\rangle : \langle \boldsymbol{\epsilon} \rangle.
    \label{eqn:Hill}
\end{equation}
The corresponding boundary conditions for the determination of the effective material properties must therefore, satisfy Eq. (\ref{eqn:Hill}). Beside others there are two boundary condition formulations that can be used for the determination of the effective material properties. Following the average strain theorem based on linear elastic material behavior as well as the average stress theorem \cite{Zohdi.2010} these boundary conditions can be written as
\begin{equation}
    \mathbf{u} = \boldsymbol{\epsilon}_0 \cdot \mathbf{x}
    \label{eqn:u}
\end{equation}
and
\begin{equation}
    \mathbf{t} = \mathbf{t}_0 \cdot \mathbf{x}.
    \label{eqn:t}
\end{equation}
Here Eq. (\ref{eqn:u}) gives the pure displacement boundary condition, where $\boldsymbol{\epsilon}_0$ is a constant macroscopic strain. In contrast to this, Eq. (\ref{eqn:t}) gives the pure traction boundary condition, with $\mathbf{t}_0$ being a constant macroscopic stress. As both boundary conditions are defined on the complete surface of the RVE, Eq. (\ref{eqn:u}) is a boundary condition of Dirichlet type and Eq. (\ref{eqn:t}) gives a boundary condition of Neumann type.

Considering a SVE on the mesoscale, the structure response depends on the boundary conditions as well as the contrast of the material components. The size of the SVE is usually described by the dimensionless parameter $\delta$, that is given by 
\begin{equation}
    \delta = \frac{d}{l}.
\end{equation}
With respect to fiber-reinforced composites the contrast is given by
\begin{equation}
    \alpha = \frac{E_{\text{f}}}{E_{\text{m}}},
\end{equation}
where $E_{\text{f}}$ represents the Young's modulus of the fiber and $E_{\text{m}}$ of the matrix material, respectively. It can be shown that the application of the presented boundary conditions on the mesoscale does not lead to identical elasticity tensors. In accordance with upper and lower bounds of effective material properties by Voigt \cite{Voigt.1889} and Reuss \cite{Reuss.1929} pure displacement boundary conditions lead to a stiffer response than the application of pure traction boundary conditions. As shown by \cite{Guan.2015} the experimentally obtained effective material properties lay between these two bounds. Therefore,
\begin{equation}
    \langle \mathbb{S}^{t}_{\delta} \rangle^{-1} \le \mathbb{C}_{\delta}^{\text{eff}} \le \langle \mathbb{C}_{\delta}^{u}\rangle
    \label{eqn:SCC}
\end{equation}
holds \cite{ Zohdi.2010,Hill.1952,OstojaStarzewski.2006}. The index $\delta$ indicates the scale dependence, whereas $u$ and $t$ represent the displacement and traction boundary conditions, respectively.

\subsection{Analytical modeling of SFRC}

\subsubsection{Mean field theory approach}
The first analytic approach for the homogenization of a microstructure was presented by Voigt \cite{Voigt.1889} and Reuss \cite{Reuss.1929} who introduced an upper and lower bound for the effective material properties. Voigt formulated these material properties based on a constant strain field, whereas Reuss used a uniform stress field. These approaches belongs to the mean-field theory. 

Many approaches that are recently used in this context are based on the work of Eshelby, who analyzed the influence of an elliptic inclusion \cite{Eshelby.1957}. This initial work was then used by Mori and Tanaka. They extended the original work adding influencing effects if more than one inclusion is added to the material \cite{Mori.1973}. Based on this framework, Tandon and Weng derived explicit formulations for the elastic constants. However, the formulation of the Poisson's ratios are coupled and therefore, needs to be determined by an iterative procedure. This was solved by Tucker and Liang \cite{TuckerIII.1999}, who found a decoupled formulation, which is also used in this study. The complete framework of the material parameter determination by Tandon and Weng is
given in \cite{Tandon.1984,TuckerIII.1999}.

\subsubsection{Self-consistent method}
A second approach for the formulation of the engineering constants for heterogeneous material is based on the self-consistent method by Hill \cite{Hill.1964}. One well-known material model in this context is given by Halpin and Tsai, who developed a semi-empiric material model for SFRC \cite{Halpin.1969,Halpin.1976}. Due to the simple implementation, this material model is widely in use \cite{Hine.2002}. In contrast to the material model by Tandon and Weng, only the Young's Modulus $E_1$ is formulated as a function of the fiber geometry. All other engineering constants remain constant for a variation of the fiber length and fiber diameter. Besides this, the fiber volume fraction influences all engineering constants. Detailed information about the determination of the engineering constants is provided in  \cite{Halpin.1969,TuckerIII.1999}.

\subsubsection{Elasticity tensor for transversely-isotropic material behavior}
There are many different possibilities to define a coordinate system in the context of fiber-reinforced materials. Commonly used is a so-called local or lamina coordinate system. Here, the first axis coincide with the fiber direction, whereas the second and third axes are perpendicular to the fiber direction. The first and second axes are in-plane with the lamina and the third axis points into the lamina thickness direction.

Based on this local coordinate system definition the elasticity tensor of transversely-isotropic material is given by
\begin{equation}
    \mathbb{C} =
    \begin{bmatrix}
    C_{11} & C_{12} & C_{13} & 0 & 0 & 0 \\
    C_{12} & C_{22} & C_{23} & 0 & 0 & 0 \\
    C_{13} & C_{23} & C_{33} & 0 & 0 & 0 \\
    0 & 0 & 0 & \frac{1}{2} (C_{22}-C_{23}) & 0 & 0\\
    0 & 0 & 0 & 0 & C_{66} & 0\\
    0 & 0 & 0 & 0 & 0 & C_{66}\\
    \end{bmatrix}.
\end{equation}
It consists of five independent coefficients, that can be calculated by five independent engineering constants $E_{1}$, $E_{2}$, $G_{12}$, $\nu_{12}$, and $\nu_{23}$. Instead of $\nu_{23}$ the out-of-plane shear modulus $G_{23}$ can be used as well. Using these engineering constants the reduced elasticity coefficients for a plane stress state can be obtained by \citep{Altenbach.2018}
\begin{eqnarray}
C_{11} &=& \frac{E_1}{(1-\nu_{21}\nu_{12})}, \label{eqn:ESZ1}\\
C_{12} &=& \frac{\nu_{21} E_1}{(1-\nu_{21}\nu_{12})} = \frac{\nu_{12} E_2}{(1-\nu_{21}\nu_{12})}, \\
C_{22} &=& \frac{E_2}{(1-\nu_{21}\nu_{12})}, \\
C_{66} &=& G_{12}. \label{eqn:ESZ2}
\end{eqnarray}
For the plane strain state the framework is more complex as the material parameters characterizing the properties in thickness direction are involved, too. The elasticity coefficients are calculated by \cite{Zienkiewicz.2002}
\begin{eqnarray}
C_{11} &=& \frac{(1-\nu_{23}^2)E_1}{(1 + \nu_{23})(1 - \nu_{23} - 2\nu_{12}\nu_{21})}, \\
C_{12} &=& \frac{\nu_{21} E_1}{(1 - \nu_{23} - 2\nu_{12}\nu_{21})}, \\
C_{22} &=& \frac{(1-\nu_{12}\nu_{21})E_2}{(1 + \nu_{23})(1 - \nu_{23} - 2\nu_{12}\nu_{21})}, \\
C_{66} &=& G_{12}.
\end{eqnarray}

The influence of a varying fiber orientation is introduced by a transformation from the local to a global coordinate system. This can be written in matrix form as
\begin{equation}
    \mathbb{C}^{'} = \mathbf{T} \mathbb{C} \mathbf{T}^{T}
    \label{eqn:transform}
\end{equation}
with the transformation matrix for a elasticity matrix reduced to a two-dimensional case
\begin{equation}
    \mathbf{T} = 
    \begin{bmatrix}
    \cos^2\theta & \sin^2\theta &  2 \cos\theta\sin\theta \\
    \sin^2\theta & \cos^2\theta &  -2 \cos\theta\sin\theta \\
    -\cos\theta\sin\theta & \cos\theta\sin\theta & \cos^2\theta-\sin^2\theta
    \end{bmatrix},
\end{equation}
where the angle between the local and global coordinate system is given by $\theta$. To analyze the influence of the fiber orientation on the engineering constants these parameters need to be extracted from the elasticity tensor defined in the global coordinate system. In case of a plane strain state this is not possible without a further assumption as there are only four independent equations containing five engineering constants. Due to the two dimensional modeling the material properties in thickness direction may be assumed constant and not affected by the fiber orientation.

\subsection{Correlated second-order random fields}
Random variables $Z$ can be used to describe quantities whose values are determined by a random experiment, which is subject to the rules of probability theory. A realization of the random variable $Z$ is given by $z$. If a random variable is furthermore, assigned to spatial coordinates, one speaks of a random field $Z(\mathbf{x})$. In the context of the continuum mechanics and for the synthesis of such random fields by using the Karhunen-Lo\`{e}ve expansion, it is necessary that the variance of the random field as well as the random variable is finite. In this case the following definition for a realization $\omega$ holds
\begin{equation}
Z(\mathbf{x}) = Z(\omega,\mathbf{x}) \in L^2(\Omega; \mathbb{R})
\end{equation}
and one speaks of second-order random fields and second-order random variables, respectively \cite{Bertein.2007}. Their main properties are briefly presented below.

Random variables are characterized by two functions. First the probability of $Z \le z$ is expressed by the cumulative distribution function
\begin{equation}
    P( Z \le z) = F_{Z}(z) = F(z).
\end{equation}
The second characteristic function is the first derivative of the cumulative distribution function called probability density function
\begin{equation}
    f(z) = \frac{\operatorname{d} F(z)}{\operatorname{d} z}.
\end{equation}
A well-known function here is the Gaussian bell curve, which represents a normal distribution.

Like random variables, random fields are characterized by a cumulative probability distribution and probability density function as well. Furthermore, both random variables and random fields are characterized by moments of probability distribution \cite{Sudret.November2000,Vanmarcke.2010}. In general the $n$-th moment of a single random variable Z is defined as
\begin{equation}
    E[Z^n] = \int_{-\infty}^{\infty} z^n f_Z(z) \operatorname{d}z.
    \label{eqn:ProMom}
\end{equation}
Based on this definition the first moment, also called expected value, is given by
\begin{equation}
    E[Z] = \mu_Z = \int_{-\infty}^{\infty} z f_Z(z) \operatorname{d}z.
\end{equation}
The second moment of a random variable is known as the mean-square of $Z$. In addition to moments of the probability distribution so called central moments can be formulated  considering the expected value. By using the second central moment the deviation of the values with respect to the expected values can be measured, which is also known as the variance. Reformulating Eq. (\ref{eqn:ProMom}) leads first to the general definition of central moments
\begin{equation}
    E[(Z-\mu_Z)^n] = \int_{-\infty}^{\infty} (z-\mu_Z)^n f_Z(z) \operatorname{d}z
    \label{eqn:ZenProMom}
\end{equation}
and therefore,
\begin{equation}
    E[(Z-\mu_Z)^2] = \int_{-\infty}^{\infty} z^2 f_Z(z) \operatorname{d}z - \mu_z^2
\end{equation}
holds for the variance of the random variable $Z$. In addition, the standard deviation is often used, which can be derived from the variance by
\begin{equation}
    \sigma_Z = \sqrt{\operatorname{Var}(Z)}.
    \label{eqn:Var}
\end{equation}
The definitions provided in Eqs. (\ref{eqn:ProMom}) to (\ref{eqn:Var}) can be easily transferred to random fields by replacing $Z$ with $Z(\mathbf{x})$. Therefore, in general the expected value as well as the variance are functions of the spatial coordinates $\mathbf{x}$. However, in case of a homogeneous random field both the expected value and the variance become constants \cite{Vanmarcke.2010}. 

The observation of a random field at different locations $\mathbf{x}_i$ is described by the corresponding random variables $\mathbf{Z} = Z_i$. In this case the relation between these random variables is expressed by the covariance, which is defined for two random variables $Z_1$ and $Z_2$ as
\begin{eqnarray}
    \operatorname{Cov}(X_1,X_2) &=& E[(X_1-\mu_1)(X_2-\mu_2)]\\
                                &=& E[X_1X_2] - \mu_1\mu_2.
                                \label{eqn:cov}
\end{eqnarray}
Usually this expression is reduced to a dimensionless parameter. Therefore, Eq. (\ref{eqn:cov}) is divided by $\sigma_1$ and $\sigma_2$, which leads to
\begin{equation}
   \rho(X_1,X_2) = \frac{\operatorname{Cov}(X_1,X_2)}{\sigma_1\sigma_2}.
   \label{eqn:DimCov}
\end{equation}
Here $\rho(X_1,X_2)$ is the dimensionless correlation parameter. If the two random variables $Z_1$ and $Z_2$ are part of the same random field $Z(\mathbf{x})$, Eqs. (\ref{eqn:cov}) and (\ref{eqn:DimCov}) give the auto-covariance and auto-correlation, respectively. In case they belong to two different random fields $Y(\mathbf{x})$ and $Z(\mathbf{x})$ the results of Eqs. (\ref{eqn:cov}) and (\ref{eqn:DimCov}) are the cross-covariance and cross-correlation, respectively. 

Usually the probability density function is unknown and hence, the random field is represented by a discrete number realizations $\omega_i$ \cite{Papoulis.1991}. In this case the mean of the discrete values
\begin{equation}
    \overline{Z(\mathbf{x})} = \frac{1}{N} \sum_{i=1}^N Z(\omega_i, \mathbf{x})
\end{equation}
can be used as expected value of the random field. In addition the variance is rewritten as
\begin{equation}
    s^2(\mathbf{x}) = \overline{Z(\mathbf{x})^2} - \overline{Z(\mathbf{x})}^2.
\end{equation}
Finally the dimensionless correlation coefficient for two random variables $Z_1$ and $Z_2$ is given by
\begin{equation}
   \rho(X_1,X_2) = \frac{\overline{[Z_1-\overline{Z_1}}][\overline{[Z_2-\overline{Z_2}]}}{s_1s_2}.
   \label{eqn:DimCovDis}
\end{equation}

\section{Analytical analysis}
In this section, the influence of geometrical fiber properties, as well as fiber volume fraction and fiber orientation on the engineering constants and the elasticity coefficients of the material, is analyzed analytically using the material models by Tandon and Weng as well as Halpin and Tsai. The goal is to identify those parameters that influence the material properties significantly and therefore, must be taken into account when analyzing the microstructural effect on the overall mechanical behavior of SFRC on a numerical basis. First, the material properties and their probabilistic characteristics are presented. This is followed by a detailed description of the analytical analysis procedure. At the end of this section, those parameters are identified that influence the mechanical properties of SFRC significantly.

\subsection{Material properties and their probabilistic characteristics}
All presented analyses are based on a polybutylene terephthalate (PBT) matrix reinforced with glass fibers. The engineering constants and densities for these two components are given in Table \ref{tab:MatPro}. For SFRC the amount of reinforcing fibers is usually expressed by the fiber mass fraction $\varphi_m$. However, within the analytical framework of material properties the fiber volume fraction $\varphi$ is used. Between these two parameters the following relation holds
\begin{equation}
    \varphi_m = \frac{\rho_f \varphi}{\rho_f \varphi + \rho_m (1- \varphi)}.
\end{equation}
In this study an overall fiber mass fraction of $\varphi_m = 30\%$ is assumed, which equals a fiber volume fraction of $\varphi = 18.22\%$.

\begin{table}[tb]
    \centering
    \begin{tabular}{llll} 
    \toprule
    & $E \ [\SI{}{\giga\pascal}]$& $\nu \ [-]$ & $\rho \  [\SI{}{\kilo\gram\per\cubic\metre}]$ \\  
    \midrule 
    Glass & $70$ & $0.22$ & $2500$\\
    PBT & $2.6$ & $0.41$ & $1300$\\ 
    \bottomrule
    \end{tabular}
    \caption{Material properties of PBT and Glass fibers.}
    \label{tab:MatPro}
\end{table}

For the investigation of the influence of SFRC characteristics like fiber length, fiber orientation, fiber diameter, and fiber volume fraction their probabilistic properties must be known. For the analytical as well as the consecutive numerical analysis in Section \ref{sec:Num} the probability distributions of these characteristics are taken from \cite{Gunzel.2013}. Like the material used in this study, the probability density functions are based on a SFRC consisting of a thermoplastic material reinforced with glass fibers and a fiber mass fraction of  $\varphi_m = 30\%$. The characteristics are determined for tensile test specimens made by mold injection.

The probability density function of the fiber length $l$ is usually approximated by a two-parameter Weibull distribution, that can be written as \cite{Chin.1988,Fu.1996}
\begin{equation}
    f(l|a,b) = \frac{b}{a} \left(\frac{l}{a}\right)^{b-1} \exp \left(-\frac{l}{a}\right)^b.
\end{equation}
For the chosen material the Weibull parameter $a$ and $b$ are set to
\begin{equation}
    a = 292 \quad b = 1.96.
\end{equation}
The corresponding mean of the fiber length is $\bar l = \SI{260}{\micro\metre}$. In contrast to the fiber length, the fiber diameter $d$ shows a normal distribution. The corresponding probability density function reads
\begin{equation}
    f(d) = \frac{1}{\sigma\sqrt{2 \pi}} \exp \left[ -\frac{1}{2} \left( \frac{d-\mu}{\sigma} \right)^2 \right]
\end{equation}
with a mean value $\mu = \SI{10.9}{\micro\metre}$ and a standard deviation $\sigma = \SI{0.9}{\micro\metre}$. The main reason why these two parameters show different probabilistic characteristics lays in the production process. While the fiber diameter is mostly influenced by the production process of the fiber itself the fiber length is not only affected by the fiber production but also by the mold injection process \cite{Gunzel.2013}.

The fiber orientation is described by an elliptic probability density function \cite{Gunzel.2013}. Therefore,
\begin{equation}
    f(\theta) = \frac{h_2}{\sqrt{1-\frac{h_1^2-h_2^2}{h_1^2}\cos^2(\theta)}}
\end{equation}
holds. Here, $h_1$ and $h_2$ are the semi-minor and semi-major axis, respectively. The semi axes ratio for this probability density function measured by $\mu$CT is 22.1. 

Based on these probability density functions the distribution of the three main characteristics, fiber length, fiber diameter, and fiber orientation are obtained. Figure \ref{fig:Weibull} gives an overview of the presented probability density functions.

The spatial distribution of the fiber volume fraction depends strongly on the window size and is therefore, not described by a probability density function \cite{Savvas.2016}.

\pgfplotsset{scaled y ticks=false}

\begin{figure}[tb]
    \tikzset{font=\small}
    \centering
    \begin{tikzpicture}
	\begin{groupplot}[group style={group size= 3 by 2},height=5.5cm,
	width=0.37\textwidth]
	\nextgroupplot[
            xlabel = {Fiber length [\SI{}{\micro\metre}]},
            ylabel = {PDF [-]},
            grid = major,
            xmin = 0,
            xmax = 1000,
            ymin = 0,
            ymax = 3/1000,
            x tick label style={inner xsep=0pt,
            /pgf/number format/.cd,%
            set thousands separator={},
            fixed},
			yticklabel style={
			/pgf/number format/fixed,
			/pgf/number format/precision=3},
			every axis plot/.append style={thick},
			xlabel style={inner ysep=0pt},         
            ylabel style={inner ysep=0pt}
            ]
            \addplot[no markers, gray] 
            table[x=x, y=y]{./Diagrammdaten/test.dat};
            
	\nextgroupplot[
            xlabel = {Fiber diameter [\SI{}{\micro\metre}]},
            grid = major,
            xmin = 7,
            xmax = 15,
            ymin = 0,
            ymax = 6/10,
			every axis plot/.append style={thick},
			xlabel style={inner ysep=0pt},         
            ylabel style={inner ysep=0pt}
            ]
            \addplot[no markers, gray] 
            table[x=x, y=y]{./Diagrammdaten/Diameter.txt};

	\nextgroupplot[
            xlabel = {Fiber orientation [\SI{}{\deg}]},
            grid = major,
            xmin = -90,
            xmax = 90,
            ymin = 0,
            ymax = 0.09,
			ytick={0,  0.03,  0.06, 0.09},
			yticklabels={$0$,  $0.03$,  $0.06$, $0.09$},
			yticklabel style={/pgf/number format/fixed},
			xticklabel style={inner xsep=0pt},
			every axis plot/.append style={thick},
			xlabel style={inner ysep=0pt},         
            ylabel style={inner ysep=0pt}
            ]
            \addplot[no markers, gray] 
            table[x=x, y=y]{./Diagrammdaten/Angle.txt};

    \end{groupplot}
    \end{tikzpicture} 
    \caption{Probability density functions.}
    \label{fig:Weibull}
\end{figure}
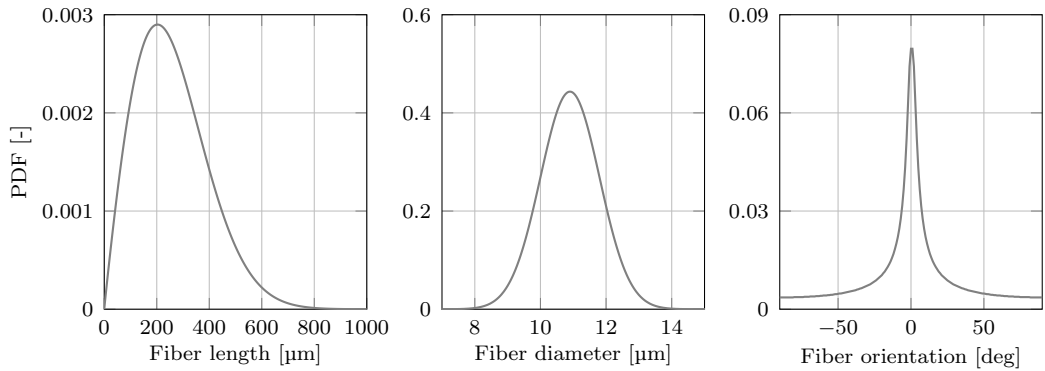

\subsection{Procedure}
For the analytic analysis, the influence of the geometrical fiber properties on the engineering constants and elasticity coefficients is investigated based on the material model by Tandon and Weng, which fits experimentally obtained values best \cite{TuckerIII.1999,Gusev.2000}. As a second approach the material model by Halpin and Tsai is used. This material model is commonly used due to the simple implementation process \cite{Hine.2002}. Both material models provide equations for the engineering constants of SFRC as a function of the fiber length, fiber diameter and the fiber volume fraction.

The influence of the different geometrical fiber properties on the elasticity coefficients is analyzed separately on an analytical basis. Therefore, the engineering constants are calculated based on the mean values given for the fiber length and fiber diameter. The chosen fiber volume fraction equals a fiber mass fraction of $30\%$. In addition one of these parameters is varied. This is done by generating 1e6 values of the varying parameter following the probability density function. For each value first the engineering constants using the material model of Tandon and Weng are calculated. Based on the material model by Halpin and Tsai only the varying Young's Modulus $E_1$ is considered. The remaining engineering constants are independent of the fiber geometry. Based on these engineering constants the elasticity coefficients are calculated assuming a plane stress state and hence, using Eqs. (\ref{eqn:ESZ1}) to (\ref{eqn:ESZ2}).

A determination of the fiber orientation influence can only be made indirectly by first determining the elasticity tensor based on the engineering constants for the local coordinate system. This elasticity tensor can then be related to a fiber orientation by a coordinate transformation using Eq. (\ref{eqn:transform}). By calculating the effective material properties based on the resulting elasticity tensor referring to a global coordinate system, it is finally possible to determine the engineering constants concerning a varying fiber orientation. All presented analyses consider a two-dimensional representation under plane stress assumption. However, analytical analyses based on a plane strain state show similar results. 

As there is no probability density function describing the fiber volume fraction, the resulting values are plotted for the range of $10\%$ to $30\%$. This fits the results of an analysis for a spatial fiber volume fraction fluctuation based on the moving window method presented in \cite{Savvas.2016}.

\subsection{Results}
First of all Table \ref{tab:MatResults} gives the engineering constants calculated for both material models using the mean values for the fiber length of \SI{260}{\micro\metre} and for the fiber diameter of \SI{10.9}{\micro\metre}. The fiber mass fraction is set to $30\%$ and the fibers are assumed to be all aligned in the $0^\circ$ direction. The values of the shear moduli and Poisson ratios are almost identically, whereas the Young's moduli show a deviation of up to $8\%$. This meets the results presented in \cite{TuckerIII.1999}.

\begin{table}[tb]
    \centering
    \begin{tabular}{llllll} 
    \toprule
    & $E_1 \ [\SI{}{\giga\pascal}]$& $E_2 \ [\SI{}{\giga\pascal}]$ & $G_{12} \ [\SI{}{\giga\pascal}]$& $G_{23} \ [\SI{}{\giga\pascal}]$ & $\nu_{12} \ [-]$  \\  
    \midrule 
    Tandon-Weng & $12.4$ & $3.99$ & $1.31$ & $1.26$ & $0.379$\\
    Halpin-Tsai & $11.2$ & $4.12$ & $1.30$ & $1.25$ & $0.375$\\ 
    \midrule 
    Deviation & $1.2$ & $0.13$ & $0.01$ & $0.01$ & $0.004$\\
    \bottomrule
    \end{tabular}
    \caption{Results of the engineering constants for Tandon-Weng and Halpin-Tsai.}
    \label{tab:MatResults}
\end{table}

\begin{figure}[p]
\centering
\includegraphics[width=0.92\textwidth]{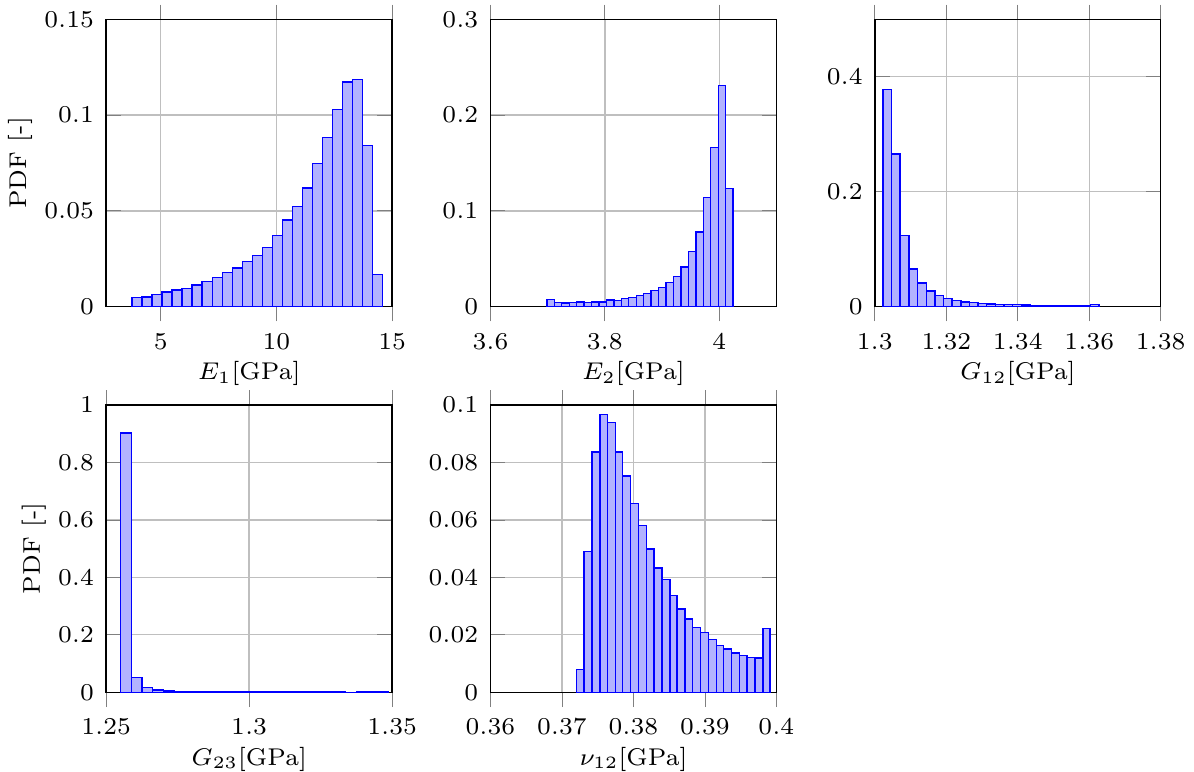}
\caption{Engineering constants due to a varying fiber length for the material model by Tandon and Weng.}
\label{fig:Tandon-Weng-ana}
\end{figure}

\begin{figure}[p]
\centering
\includegraphics[width=0.92\textwidth]{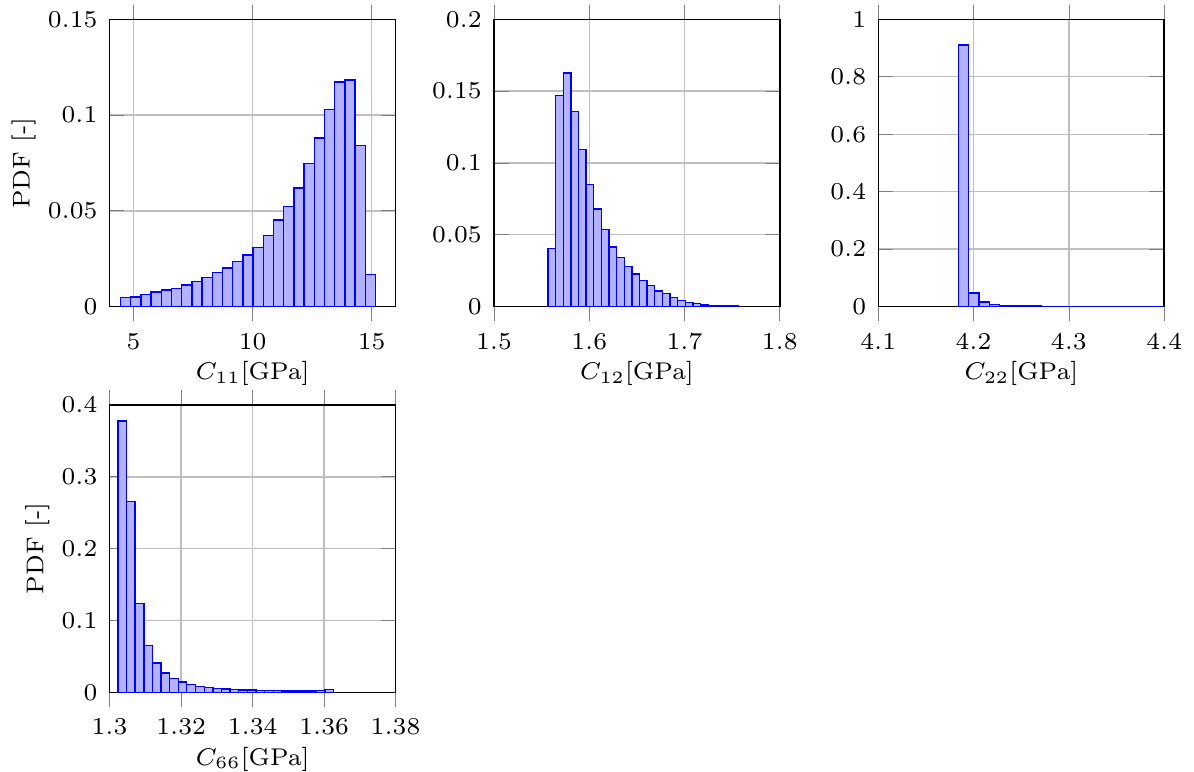}
\caption{Elasticity coefficients due to a varying fiber length for the material model by Tandon and Weng.}
\label{fig:Tandon-Weng-ana-c}
\end{figure}

As an example, Figure \ref{fig:Tandon-Weng-ana} shows the results of the engineering constants based on the material model by Tandon and Weng with a varying fiber length. It can be seen that the fiber length influences the Young's modulus $E_1$ significantly, whereas the shear moduli are almost independent. The Young's modulus $E_2$ as well as the Poisson ratio $\nu_{12}$ react only slightly to a variation of the fiber length. Following this only the coefficient $C_{11}$ is significantly influenced by the fiber length. A minor influence can be recognized for the coefficient $C_{12}$ as this coefficients depends also on $E_1$. The remaining coefficients $C_{22}$ and $C_{66}$ are nearly independent (see Figure \ref{fig:Tandon-Weng-ana-c}). Furthermore, the distribution of the engineering constants as well as the elasticity coefficients can be approximated by a Weibull distribution, which also describes the distribution of the fiber length itself.

\begin{figure}[p]
\centering
\includegraphics[width=0.92\textwidth]{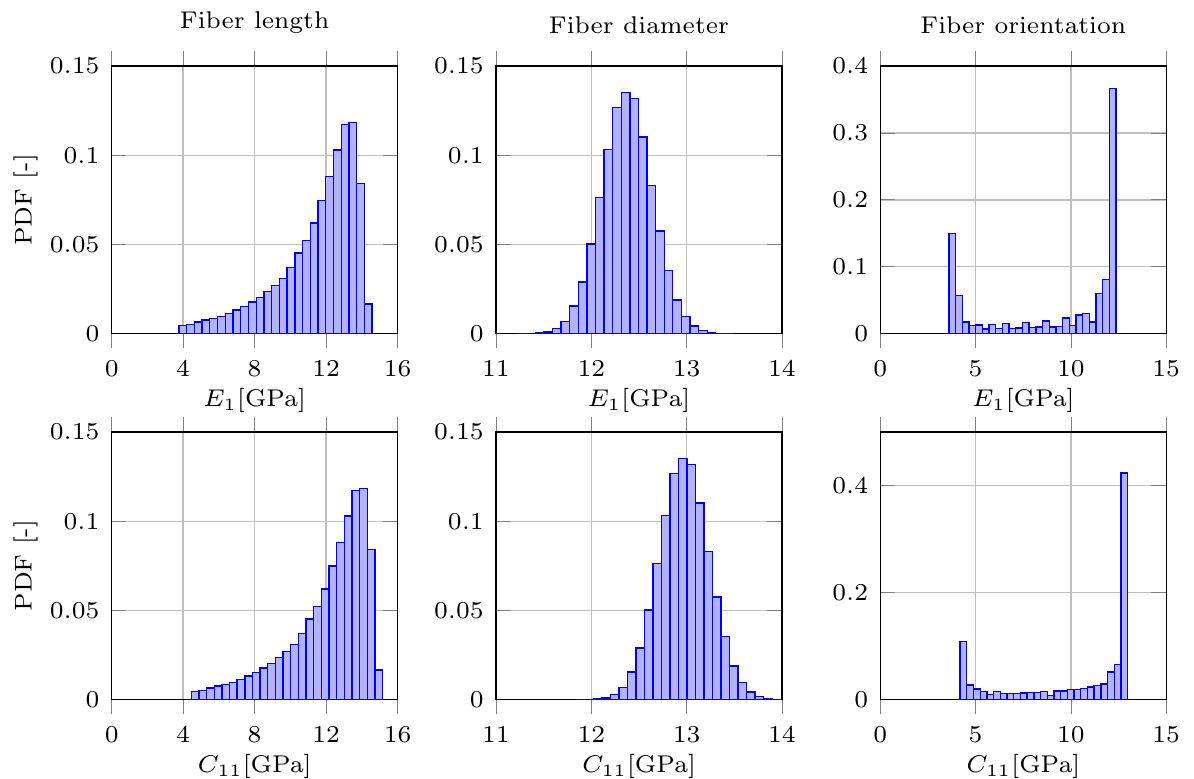}
\caption{Results for the distribution of $E_1$ and $C_{11}$ with respect to a varying fiber length, diameter, and orientation based on the material model by Tandon and Weng.}
\label{fig:Analytic_TW}
\end{figure}

\begin{figure}[p]
\centering
\includegraphics[width=0.92\textwidth]{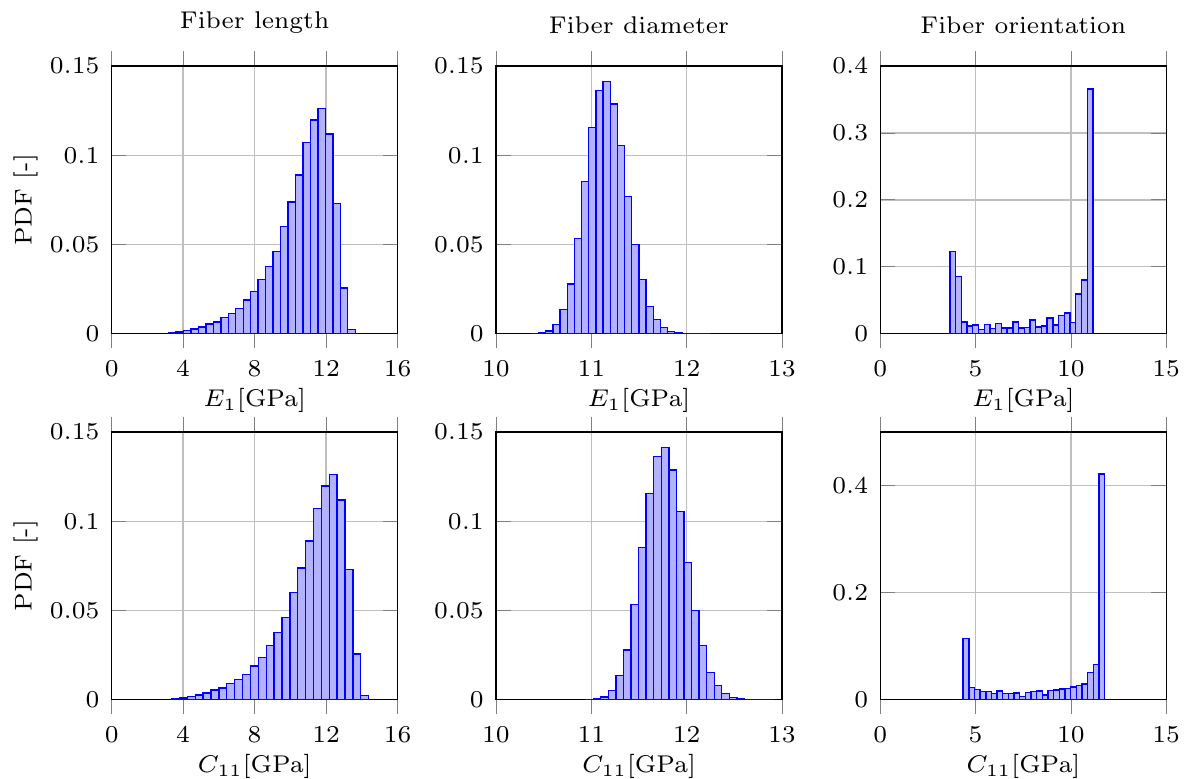}
\caption{Results for the distribution of $E_1$ and $C_{11}$ with respect to a varying fiber length, diameter, and orientation based on the material model by Halpin and Tsai.}
\label{fig:Analystic_HT}
\end{figure}

The analysis based on the material model by Halpin and Tsai leads to identical conclusions. As only $E_1$ is formulated as a function of the fiber length, the remaining constants are independent and only $C_{11}$ is significantly influenced by the fiber length. Again, the distribution of $E_1$ as well as the resulting elasticity coefficients can be approximated by a Weibull distribution.

Based on these results the following analyses of the fiber diameter influence, as well as the fiber orientation, are concentrated on $E_1$ and $C_{11}$. The full results of the analytic analysis for both material models regarding the fiber diameter and fiber orientation can be found in \ref{sec:Results_ana}. In Figures \ref{fig:Analytic_TW} and \ref{fig:Analystic_HT} the results of $E_1$ and $C_{11}$ are given based on the material model by Tandon and Weng and Halpin and Tsai, respectively. Starting from left to right they show the results for the fiber length, fiber diameter, and fiber orientation, respectively. It can be observed, that both material models lead to similar results. The engineering constants as well as the elasticity coefficients are significantly influenced by the fiber length and fiber orientation. However, the fiber diameter has just little impact on the material properties. Furthermore, the fiber orientation shows very high values for angles close to zero. As the fiber angle increases the values are dropping rapidly. This is indicated by a second local maximum at the lower spectrum of the values, which corresponds to fiber orientation of $60^\circ$ up to $90^\circ$. 

\begin{figure}[t]
\centering
\includegraphics[width=\textwidth]{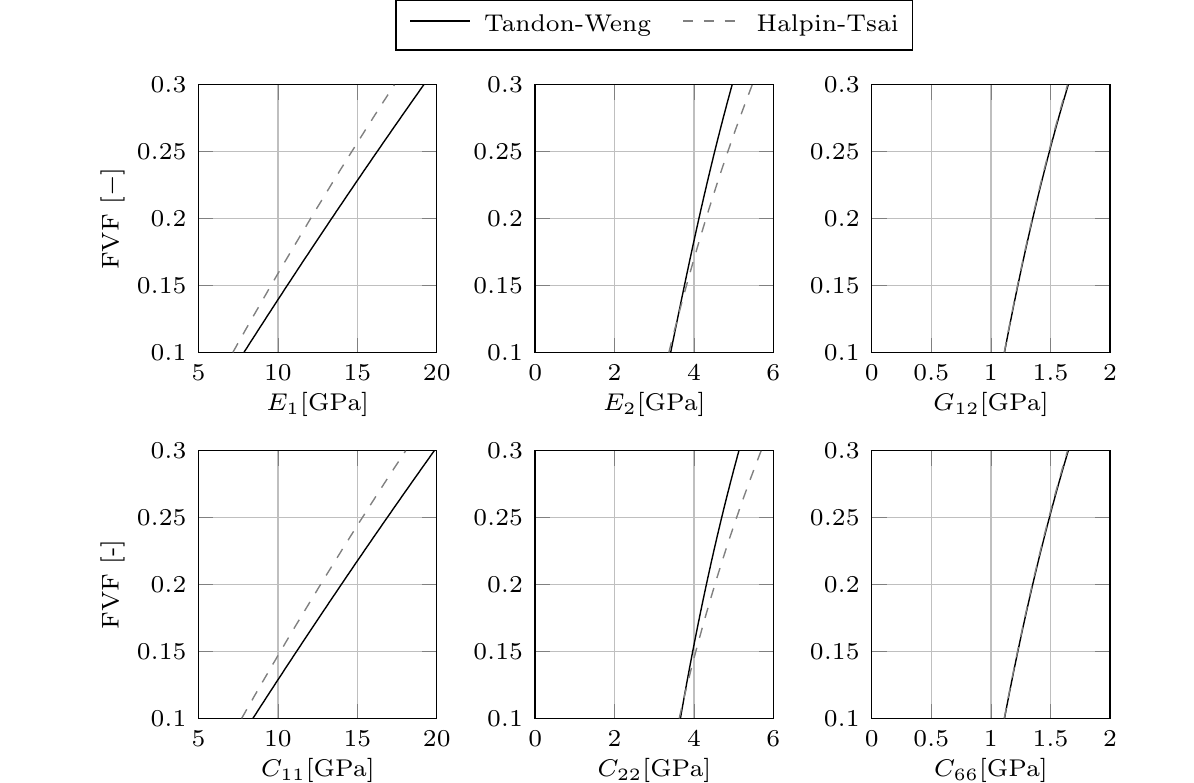}
\caption{Influence of the fiber volume fraction on the material properties of SFRC.}
\label{fig:FVF}
\end{figure}

Finally Figure \ref{fig:FVF} gives the results of main material properties as a function of the fiber volume fraction for both material models. As shown in \cite{Savvas.2016} the fiber volume fraction influences significantly the parameters $E_1$ and $C_{11}$. Furthermore, the influence on $E_2$ and $G_{12}$ and the corresponding elements of the elasticity tensor is still significant for both material models in the range of $10\%$ to $30\%$ fiber volume fraction.

Based on the results of this analytical analysis it can be concluded that the fiber length as well as the fiber orientation and fiber volume fraction influence the material properties of SFRC significantly, whereas the fiber diameter shows only a minimal influence. This corresponds to the results presented in \cite{PflammJonas.2001}. Furthermore, it can be observed that the distribution of the engineering constants, as well as the elasticity coefficients, can be approximated with the same type of probability density function as the varied parameter itself. This also holds when assuming a plane strain state.

\section{Numerical analysis}
\label{sec:Num}

The numerical analysis covers the influence of the microstructure properties of SFRC on the mechanical behavior as well as the correlation structure of the resulting elasticity tensor. First, the numerical procedure is presented. This is followed by the determination of a sufficient element size for the numerical model. Finally, the influence of the microstructure on the mechanical properties is analyzed and the correlation structure of the elasticity tensor is presented.

\subsection{Generation of the microstructure}

\begin{figure}[t]
    \centering
    \includegraphics[width=0.95\textwidth]{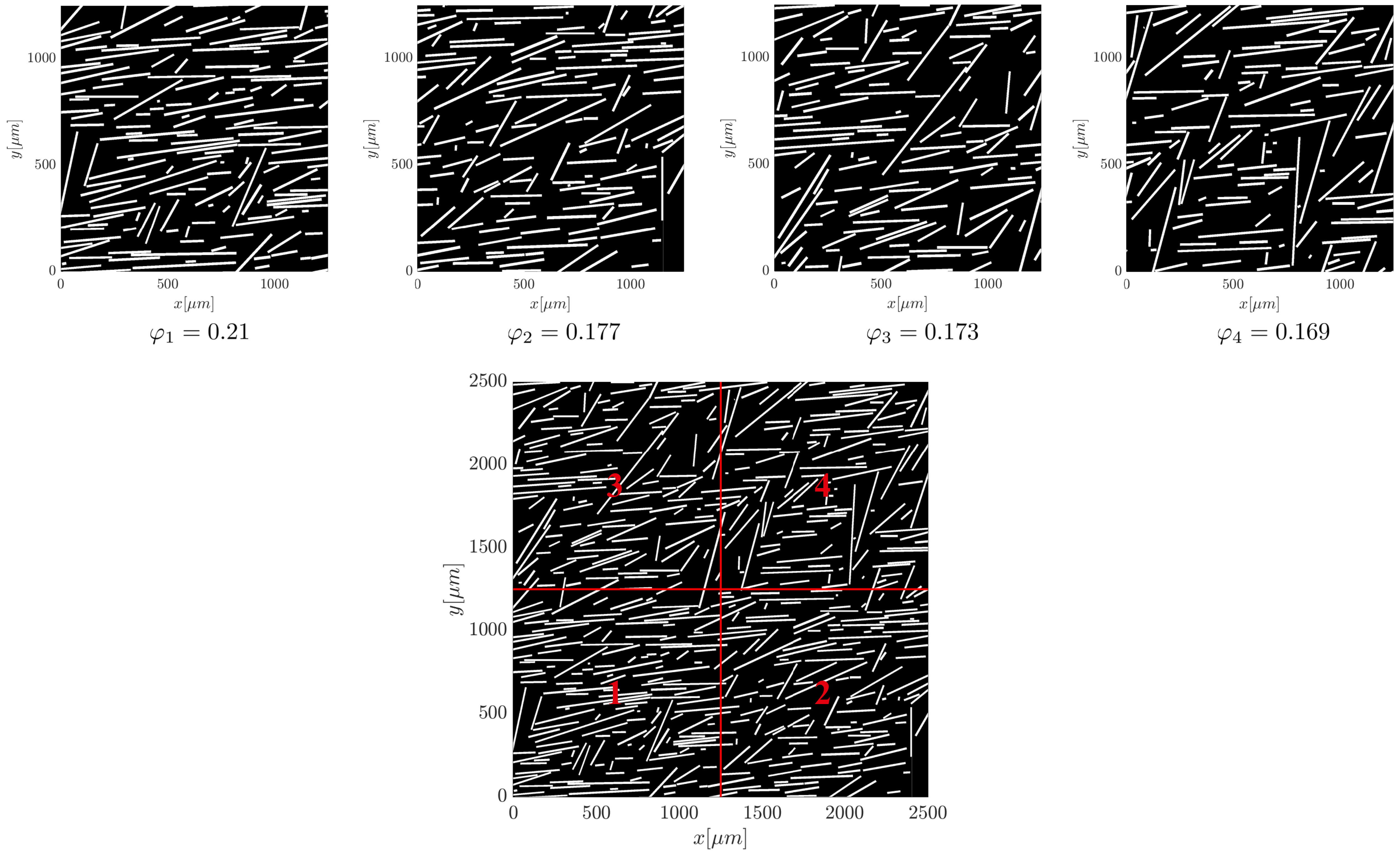}
    \caption{SFRC microstructure of $\SI{2500}{\micro\metre} \times \SI{2500}{\micro\metre}$ with a over all fiber volume fraction of $18.2\%$.}
    \label{fig:Microstr}
\end{figure}

For all numerical analyses first, a microstructural model is generated representing randomly placed fibers in a preset area. Later this microstructure is transferred to a numerical model. The generation of the microstructure is based on an adapted Poisson process. First, a set of two integers within the preset area is randomly chosen. These integers represent the midpoint of the new fiber. Next, the fiber length, fiber diameter and the fiber orientation of this fiber are determined following their probability density functions. With all the necessary information about the fiber, it is checked if there are any overlaps with other fibers. Only if there are no overlaps the fiber is finally added to the area, which is represented by a two-dimensional array with a step size of \SI{1}{\micro\metre} along both axes. In comparison with an average fiber dia\-me\-ter of \SI{10}{\micro\metre} and an average fiber length of \SI{260}{\micro\metre} the grid size of \SI{1}{\micro\metre} is sufficient. The procedure is repeated until a predefined fiber volume fraction is reached. Figure \ref{fig:Microstr} shows an exemplarily generated microstructure with a size of $\SI{2500}{\micro\metre} \times \SI{2500}{\micro\metre}$. Here, a significant difference between the analytical prediction and the numerical modeling is getting obvious. In contrast to the analytical description, the numerical model can consist of several fibers with different geometrical properties. Furthermore, the fiber orientation, as well as the fiber volume fraction, can show spatial fluctuations. Regarding the analytical modeling all fibers are assumed to have identical geometrical properties as well as orientations.

Below microstructures with different varying characteristics are analyzed numerically. In comparison to the analytical modeling first besides one parameter, all remaining are set to the mean value. By doing so the influence of a varying fiber length, fiber diameter, and fiber orientation can be investigated separately. In a further step, a microstructure is generated, that is much larger than the analyzed window size. Extracting a smaller window from a larger microstructure enables one, to analyze a varying fiber volume content, as depicted in Figure \ref{fig:Microstr}. In a final step, a microstructure is generating where all characteristics are set with respect to their probabilistic characteristic.

\subsection{Finite Element Model}
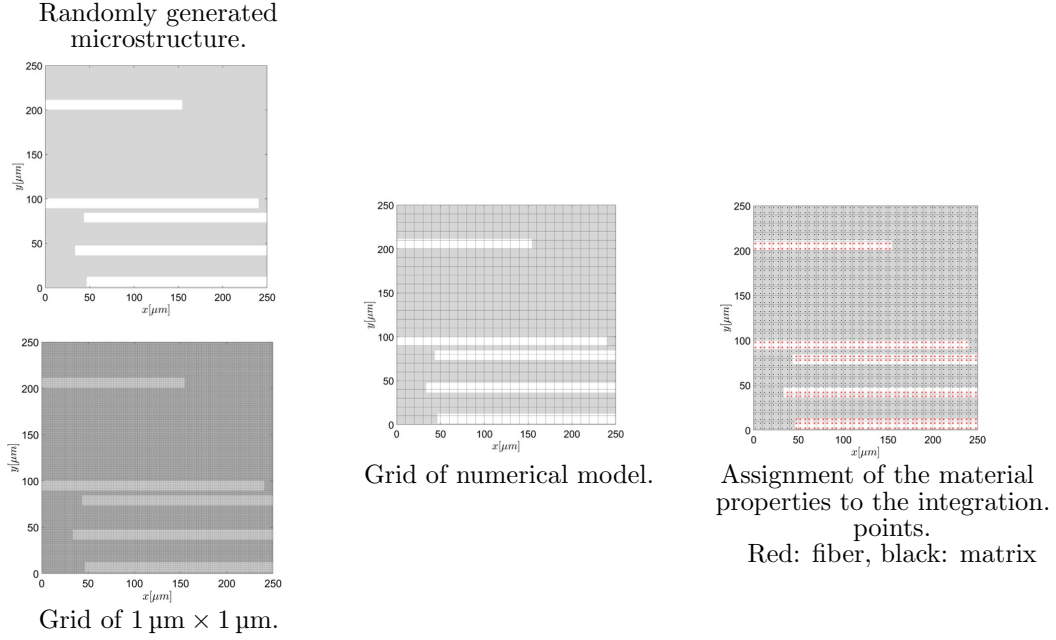
\begin{figure}[p]
\begin{picture}(320,250)
\put(0,15){\includegraphics[width=0.95\textwidth]{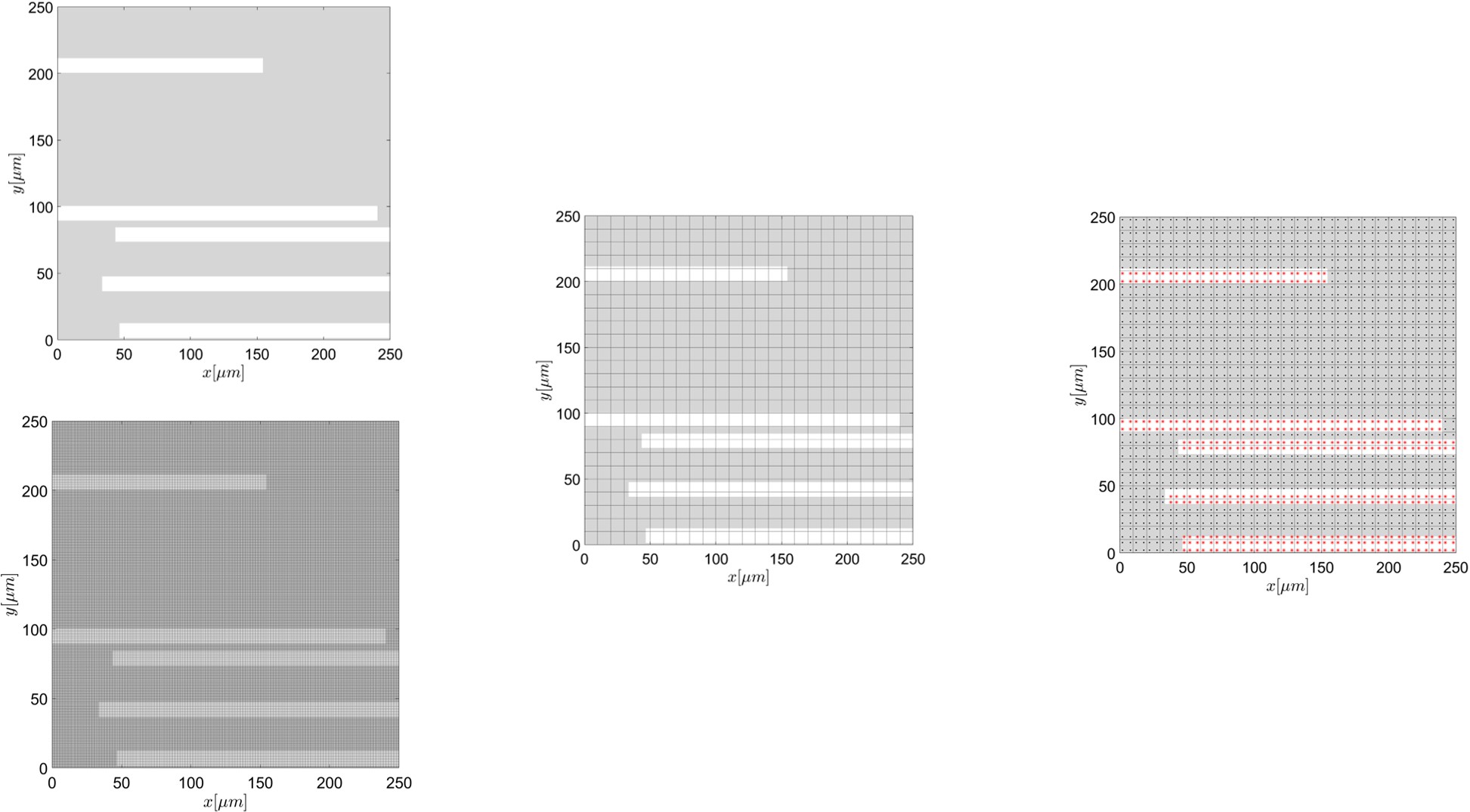}}
\put(12,5){\footnotesize{Grid of $\SI{1}{\micro\metre} \times \SI{1}{\micro\metre}$. }}
\put(12,235){\footnotesize{Randomly generated}}
\put(24,225){\footnotesize{microstructure.}}
\put(135,60){\footnotesize{Grid of numerical model.}}
\put(270,60){\footnotesize{Assignment of the material}}
\put(268,50){\footnotesize{properties to the integration.}}
\put(320,40){\footnotesize{points.}}
\put(280,30){\footnotesize{Red: fiber, black: matrix}}
\end{picture}
\caption{Procedure to generate a numerical model representing a SFRC microstructure.}
\label{fig:Procedure}
\end{figure}

After implementing the microstructure a numerical model is generated using Comsol. The numerical model consists of a square in plane stress state discretized by a structured mesh consisting of squared Lagrange elements with quadratic shape functions. In the last step, the material properties (see Table \ref{tab:MatPro}) are passed to the numerical model. This is done by first saving the material properties in arrays that have the same structure as the array representing the microstructure itself. As both components show isotropic material behavior, these arrays provide the distribution of the Young's Modulus as well as the Poisson's ratio of the two materials over the microstructure. Finally, to each integration point the corresponding material properties are passed. This procedure is depicted in Figure \ref{fig:Procedure}. However, assigning the material properties to the integration points leads to a mesh dependent representation of the microstructure. Therefore, to ensure a sufficient representation of the resulting material properties, the influence of the element size is analyzed in detail in Section \ref{sec:ele}.

\begin{table}[p]
    \centering
    \begin{tabular}{llllll} 
    \toprule
    BC & Load case & Left edge & Right edge & Upper edge & Lower edge \\
    & & $x=0$ & $x=d$ & $y=0$ & $y=d$\\
    \midrule
    & \multirow{2}*{1} & $u_1=0$ & $u_1=u_0$  & $u_1=u_0 x$ & $u_1=u_0 x$ \\
    &  & $u_2=0$ & $u_2=0$  & $u_2=0$ & $u_2=0$ \\
   Pure & \multirow{2}*{2} & $u_1=0$ & $u_1=0$  & $u_1=0$ & $u_1=0$ \\
   displ.&  & $u_2=u_0 y$ & $u_2=u_0 y$  & $u_2=u_0$ & $u_2=0$ \\
    & \multirow{2}*{3} & $u_1=0$ & $u_1=0$  & $u_1=0$ & $u_1=0$ \\
    &  & $u_2=0$ & $u_2=u_0$  & $u_2=u_0 x$ & $u_2=u_0 x$ \\
    \midrule
    & \multirow{2}*{1} & $t_1=\frac{t_0}{2}$ & $t_1=-\frac{t_0}{2}$  & $t_1=0$ & $t_1=0$ \\
    &   & $t_2=0$ & $t_2=0$  & $t_2=0$ & $t_2=0$ \\
    Pure& \multirow{2}*{2} & $t_1=0$ & $t_1=0$  & $t_1=0$ & $t_1=0$ \\
    trac.&   & $t_2=0$ & $t_2=0$  & $t_2=\frac{t_0}{2}$ & $t_2=-\frac{t_0}{2}$ \\
    & \multirow{2}*{3} & $t_1=0$ & $t_1=0$  & $t_1=\frac{t_0}{2}$ & $t_1=-\frac{t_0}{2}$ \\
    &   & $t_2=-\frac{t_0}{2}$ & $t_2=\frac{t_0}{2}$  & $t_2=0$ & $t_2=0$ \\
    \bottomrule
    \end{tabular}
    \caption{Load cases for the determination of the elasticity coefficients in accordance with the Hill condition \cite{Zimmermann.2019}.}
    \label{tab:BC}
\end{table}

To determine the elasticity coefficients the boundary conditions are defined in accordance with Eqs. (\ref{eqn:u}) and (\ref{eqn:t}). Usually, individual simulations for each elasticity tensor component are performed \cite{Zohdi.2010,OstojaStarzewski.2006}. However, in \cite{Zimmermann.2019} it is shown that it is also possible to use just three independent load cases to be able to calculate all nine elasticity coefficients of a two-dimensional model individually. This is done by formulating three independent boundary conditions, where always just one strain component is not equal zero. Table \ref{tab:BC} gives an overview of all load cases used to determine the elasticity coefficients for pure kinematic as well as pure traction boundary conditions in accordance with the Hill condition given by Eq. [\ref{eqn:Hill}]. 

\subsection{Analysis of the element size}
\label{sec:ele}
One crucial aspect of this approach is the element size. Hence, before the influence of the microstructure is analyzed by numerical simulations, a sufficient element size must be determined. This is done by  performing numerical simulations based on a microstructure of $\SI{250}{\micro\metre} \times \SI{250}{\micro\metre}$ with different element sizes. For each configuration, the pure displacement and pure traction boundary conditions are applied and the elasticity coefficients are determined based on the framework given in \ref{sec:Calc_E}. For statistical reasons, the simulation is carried out for 500 different microstructures \cite{Sena.2013,Soize.2008}. Table \ref{tab:Element_Size} provides the mean values of the elasticity coefficients first for pure displacement and afterward for pure traction boundary conditions.  

First of all the values indicate, that the symmetry of the elasticity tensor is still valid. However, there is a small deviation between the elasticity coefficients $C_{12}$ and $C_{21}$. Furthermore, the dependence on the boundary conditions of the results and hence the scale dependence of $C_{11}$, $C_{22}$ and $C_{66}$ is clearly observable. As presented in Section \ref{sec:Theory} the displacement boundary conditions lead to an upper bound whereas the traction boundary conditions lead to a lower bound. Therefore, these results confirm Eqn. (\ref{eqn:SCC}).

Comparing the values for the different element sizes with each other show only minor variations for an element size of \SI{10}{\micro\metre} or less. In this case the obtained results for the elasticity coefficients of a window with a size of $\SI{250}{\micro\metre} \times \SI{250}{\micro\metre}$ does not depend on the element size. Therefore, for all following simulations the element size is set to $\SI{10}{\micro\metre} \times \SI{10}{\micro\metre}$. 

\begin{table}[tb]
    \centering
    \begin{tabular}{llllllll} 
    \toprule
    & Element  & $C_{11}$ & $C_{12}$ & $C_{21}$ & $C_{22}$ & $C_{66}$ & $\varphi_m $\\
    & size $[\si{\micro\metre}]$& $[\SI{}{\giga\pascal}]$ & $[\SI{}{\giga\pascal}]$ & $[\SI{}{\giga\pascal}]$ & $[\SI{}{\giga\pascal}]$ & $[\SI{}{\giga\pascal}]$ & $[-]$\\
    \midrule
    \multirow{4}*{KUBC}& 16.7 & 12.0 & 1.62 & 1.51 & 4.67 & 1.47 & 0.302\\
    & 10 & 11.7 & 1.56 & 1.49 & 4.36 & 1.35 & 0.303\\
    & 5 & 11.6 & 1.53 & 1.48 & 4.21 & 1.30 & 0.302\\
    & 2.5 & 11.6 & 1.51 & 1.48 & 4.06 & 1.26 & 0.302\\
    \midrule
    \multirow{4}*{SUBC}& 16.7 & 6.34 & 1.49 & 1.60 & 4.01 & 1.20 & 0.302\\
    & 10 & 6.42 & 1.49 & 1.58 & 3.91 & 1.16 & 0.303\\
    & 5 & 6.46 & 1.51 & 1.56 & 3.89 & 1.15 & 0.302\\
    & 2.5 & 6.51 & 1.52 & 1.54 & 3.83 & 1.14 & 0.302\\
    \bottomrule
    \end{tabular}
    \caption{Results of the elasticity coefficients with respect to the element size.}
    \label{tab:Element_Size}
\end{table}

\subsection{Influence of the microstructure}

\subsubsection{Overall properties of the elasticity tensor}
\label{sec:Overall}

\begin{figure}[t]
\centering
\includegraphics[width=0.92\textwidth]{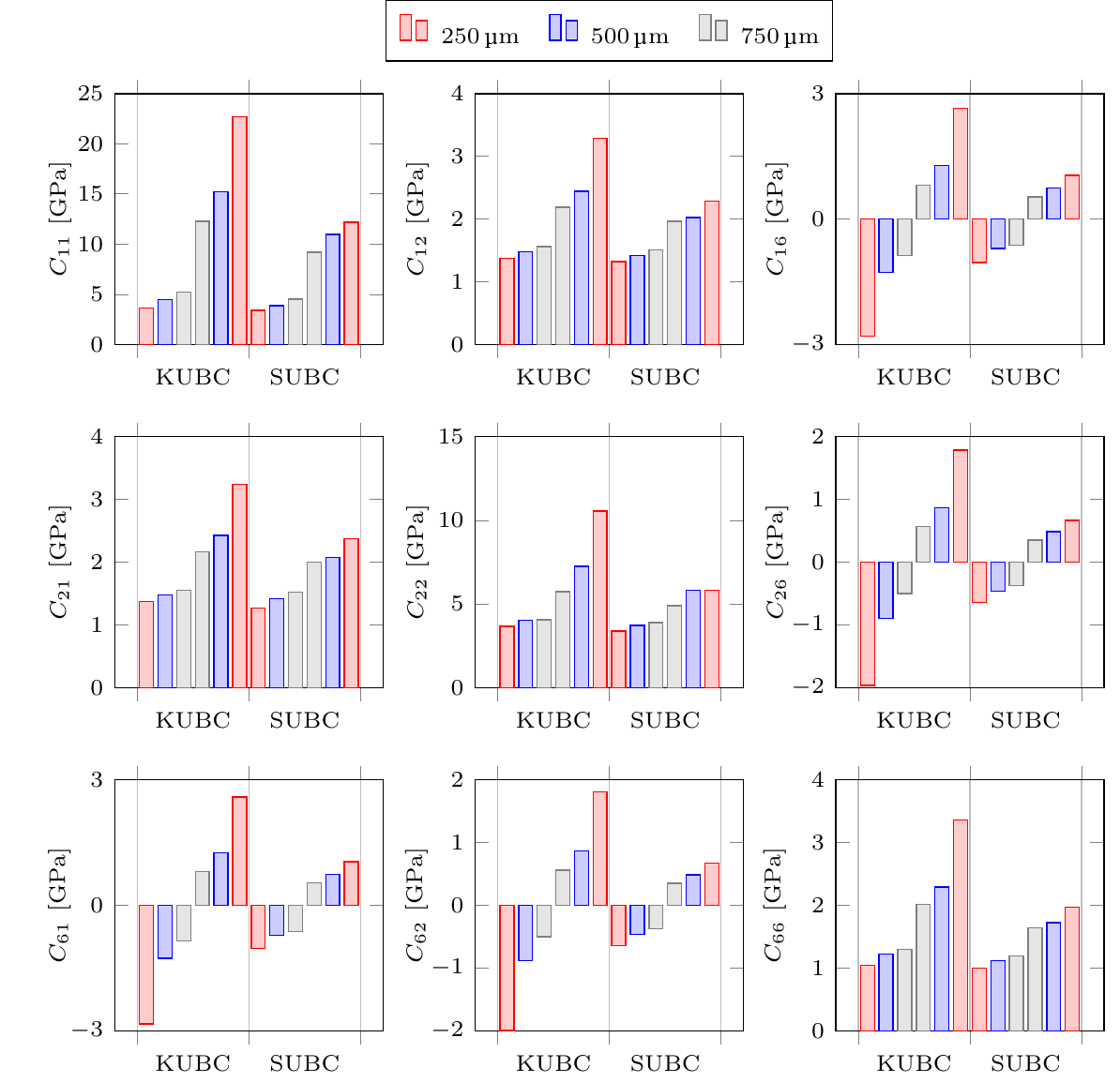}
\caption{Symmetry analysis of the elasticity tensor.}
\label{fig:auto1}
\end{figure}

Before analyzing the influencing geometrical properties in detail the overall characteristics of the elasticity tensor with respect to the window size and the boundary condition is presented. This is done based on a microstructure where all main properties are assumed to show probabilistic behavior. This ensures an overall realistic representation of the microstructure. 

Figure \ref{fig:auto1} gives the minimal and maximal value of each elasticity coefficient for both types of boundary conditions and a window size of \SI{250}{\micro\metre}, \SI{500}{\micro\metre}, and \SI{750}{\micro\metre}. First of all the scale dependence of the elasticity coefficients can be found as the range of the coefficients decreases with increasing window size. In addition, the symmetry properties of the elasticity tensor are observable. Not only $C_{12}$ and $C_{21}$ show an almost identical behavior, but also $C_{16}$ and $C_{61}$ as well as $C_{26}$ and $C_{62}$ coincide very well. However, in contrast to $C_{12}$ the mean values of $C_{16}$ and $C_{26}$ are approximately zero. As their fluctuation starts to vanish with an increasing window size the overall assumption of transversely-isotropic material behavior holds. This is also independent of the boundary condition. The overall transversely-isotropic material behavior is also indicated by the elements $C_{11}$ and $C_{22}$. Due to the aligned characteristic of the fibers predicted by the corresponding PDF, see Figure \ref{fig:Weibull}, the value of $C_{11}$ is greater than the value of $C_{22}$.

\begin{figure}
    \centering
    \includegraphics[width= \textwidth]{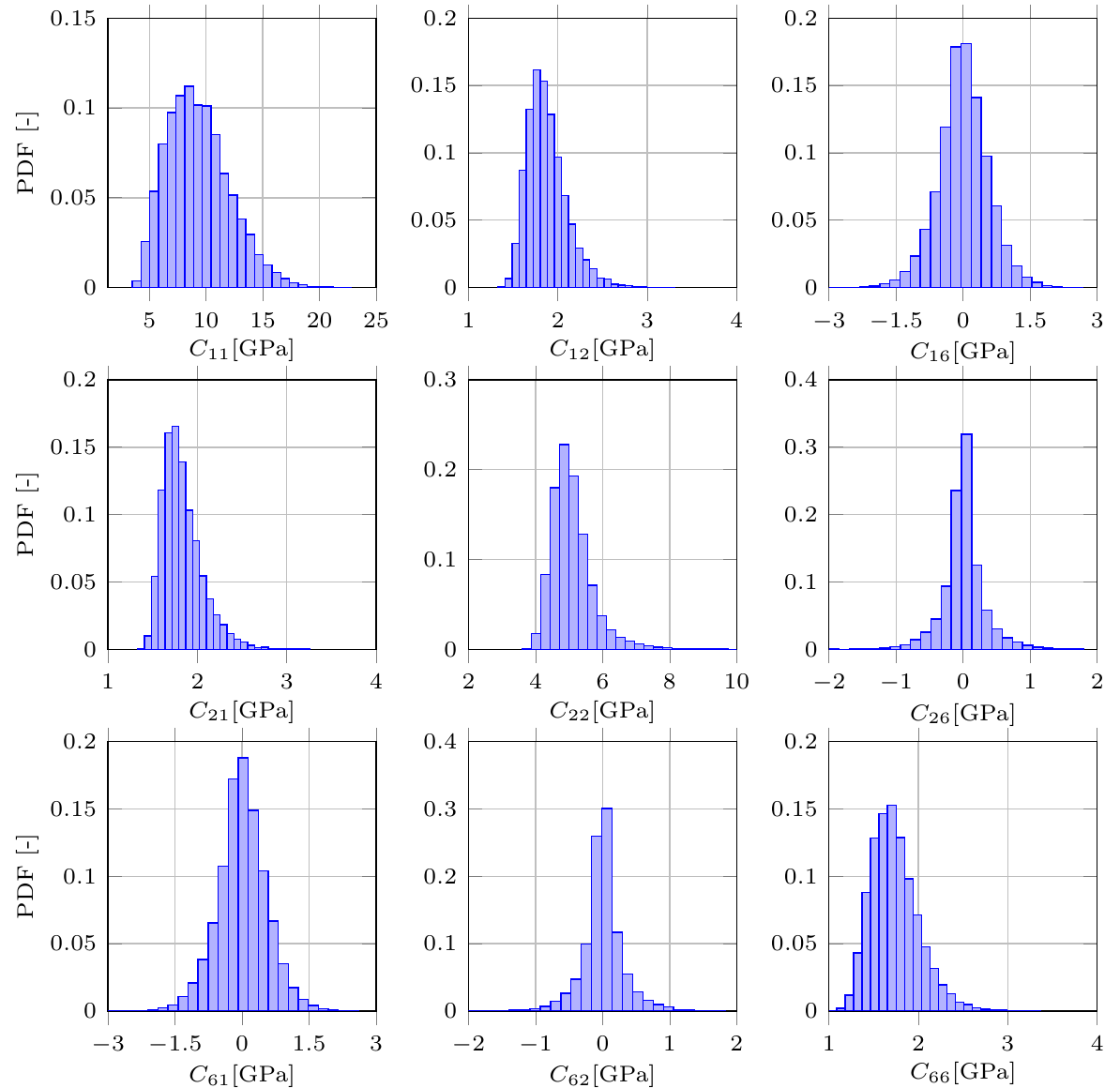}
    \caption{Distribution of the elasticity coefficients under pure displacement boundary conditions with respect to a \SI{250}{\micro\metre} window size.}
    \label{fig:250_c}
\end{figure}

Finally, the results of the elasticity coefficients show that the values for pure kinematic boundary conditions are higher compared to pure traction boundary conditions, which meets the theoretical framework of the multiscale modeling.

\begin{figure}
    \centering
    \includegraphics[width=0.9\textwidth]{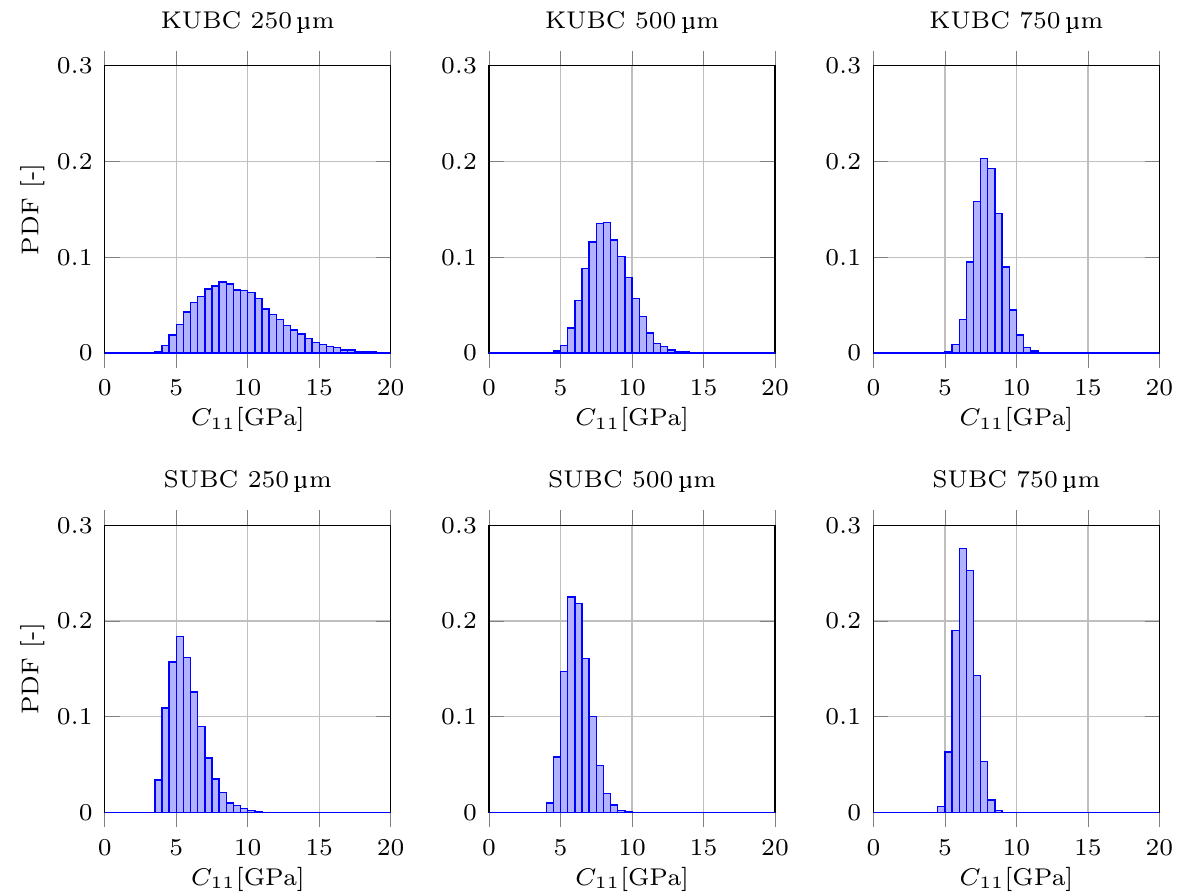}
    \caption{Distribution of the values for the elasticity tensor element $C_{11}$ with respect to the boundary conditions and the window size.}
    \label{fig:C11}
\end{figure}

For taking a closer look at the distribution of the individual coefficients Figure \ref{fig:250_c} shows the histogram of each elasticity coefficient for a window size of \SI{250}{\micro\metre} and pure displacement boundary conditions based on 16.500 data sets. Again the symmetry properties of the elasticity tensor, as well as the anisotropic effect due to the limited window size, can be observed. However, not all coefficients can be approximated best with the same probability distribution type. Whereas $C_{11}$, $C_{12}$, $C_{22}$, and $C_{66}$ seem to meet a Weibull distribution best, the remaining parameters are most likely normal distributed. Observing this over an increasing window size reveals that the distribution of the elasticity coefficients depends on the window size of the analyzed microstructure as depicted in Figure \ref{fig:C11}. In this figure the histogram of $C_{11}$ is given for both boundary condition types as well as all window sizes. The histogram indicates, that the distribution of $C_{11}$ equals a Weibull distribution for small window sizes whereas for an increasing window size the distribution starts to meet a normal distribution. This observation is independent of the boundary condition. Therefore, it can be concluded that the distribution of the elasticity coefficients is significantly influenced by the microstructural properties on the mesoscale. However, when the SVE gets close to the RVE the microstructure does not affect the distribution any longer. 

\subsubsection{Parameter identification}

For the analysis of the geometrical parameters influencing the material properties of SFRC on the mesoscale, first, the different parameters are observed individually. Finally, all parameters are combined. For the analysis of the fiber length, orientation, and diameter influence, the microstructure is generated with respect to the varying parameter and the window size. The remaining parameters are set to the mean value according to their probabilistic characteristics. Considering a locally distributed fiber volume fraction a microstructure is generated that is larger than the actual window size. By extraction a window from a greater microstructure, the corresponding fiber volume fraction is not constant, see Figure \ref{fig:Microstr}.

Again for statistical reasons, the simulations for each analysis are carried out for 500 different microstructures \cite{Sena.2013,Soize.2008}. Furthermore, as the material properties on the mesoscale depend on the window size the procedure is done for a window size of \SI{250}{\micro\metre}, \SI{500}{\micro\metre}, and  \SI{750}{\micro\metre}. 

\begin{figure}[p]
    \centering
    \includegraphics[width= \textwidth]{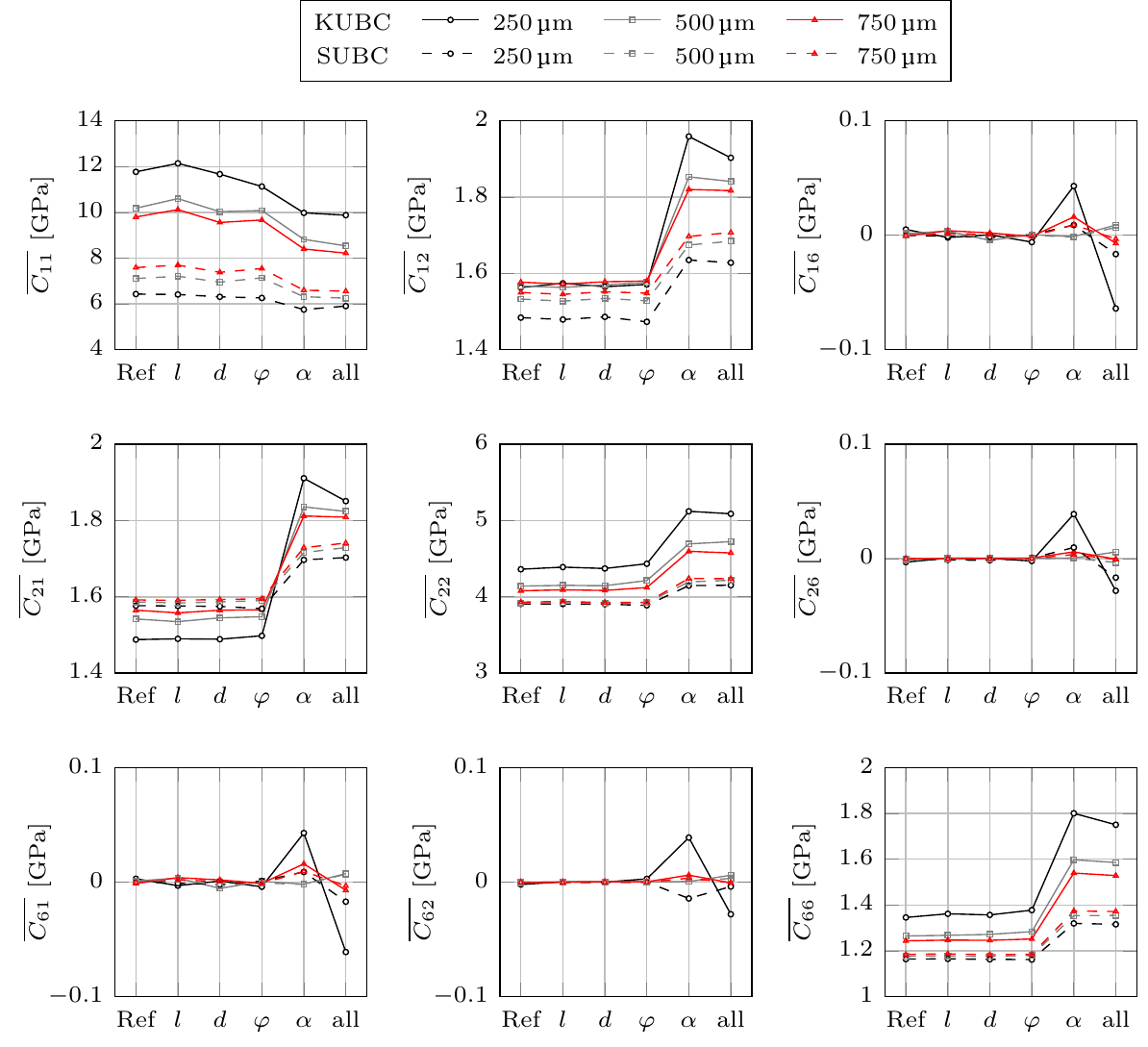}
    \caption{Mean values of the elasticity coefficients in dependence of the geometrical fiber properties as well as the fiber volume fraction and fiber orientation on the elasticity coefficients.}
    \label{fig:Influ}
\end{figure}

\begin{figure}[p]
    \centering
    \includegraphics[width= \textwidth]{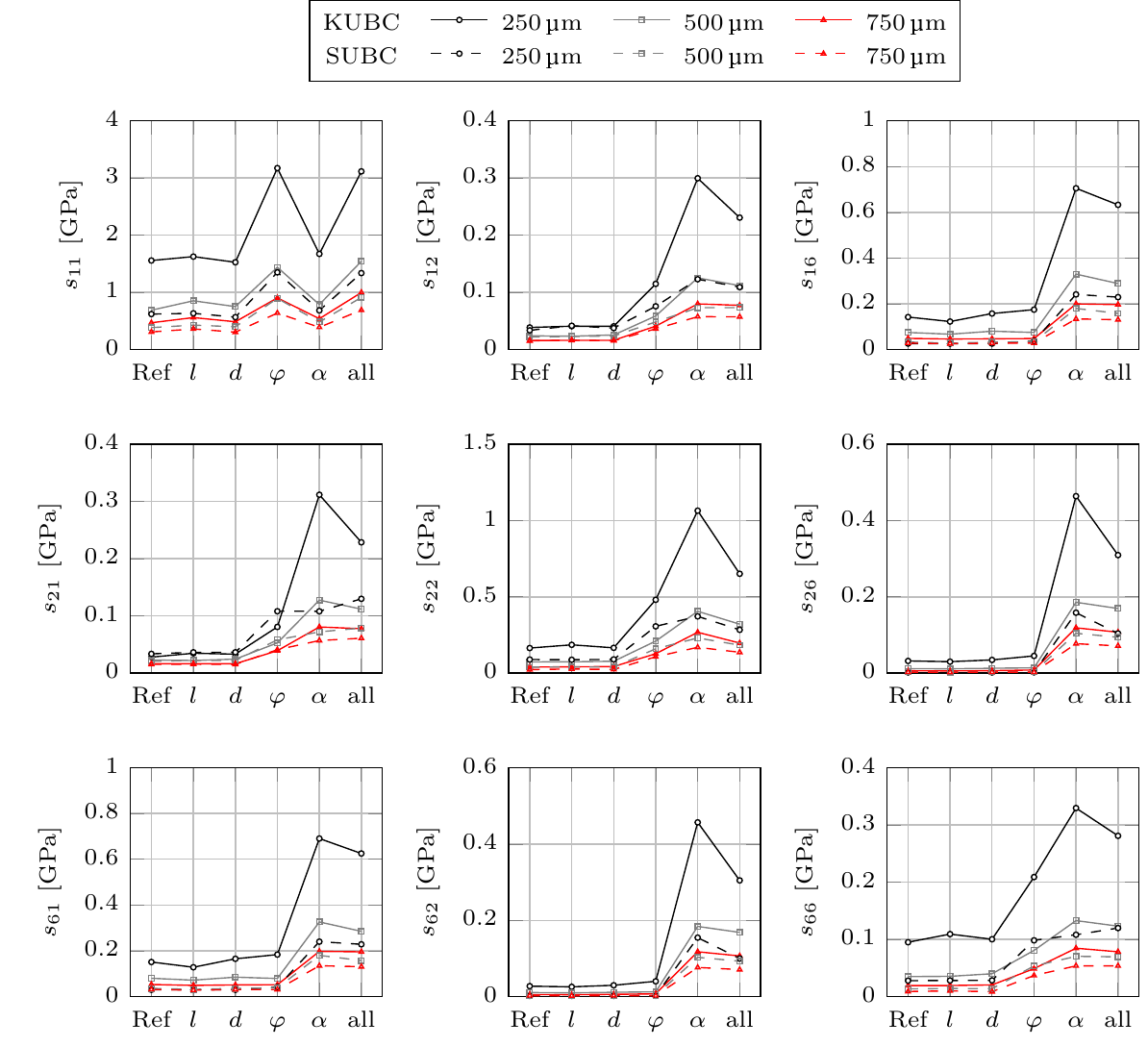}
    \caption{Standard deviation of the elasticity coefficients in dependence of the geometrical fiber properties as well as the fiber volume fraction and fiber orientation on the elasticity coefficients.}
    \label{fig:Influ3}
\end{figure}

Figure \ref{fig:Influ} and \ref{fig:Influ3} show the mean value and the standard deviation of each elasticity coefficient with respect to the boundary conditions as well as the window size for each analyzed configuration. The coefficients can be divided into two groups, those who have a mean value of zero ($C_{16}, C_{26}, C_{61}, C_{62}$) and the remaining coefficients. The coefficients with a mean value of zero show a significant scattering mainly induced by a varying fiber orientation. This is not only indicated by the standard deviation but also by the mean values of these elasticity coefficients. This fits the analytic description of composite materials. As long as the fibers are all aligned with the symmetry axis 
\begin{equation}
C_{16} = C_{26} = C_{61} = C_{62} = 0
\end{equation}
holds. However, when rotating the fiber orientation by a coordinate transform these elements are no longer equal to zero.

The sensitivity to the orientation of the fiber can also be found for the remaining elasticity coefficients. Here again, not only the mean values are significantly influenced, but also the standard deviation increases significantly. Furthermore, the standard deviation of these coefficients is also sensitive to the fiber volume fraction. Finally, there is only a slight influence of the fiber length as well as the fiber diameter.

\subsection{Comparison of the analytical and numerical results}

First of all the numerical results of the reference elasticity tensor are compared to the analytical results of the elasticity tensor based on the engineering constants given in Table \ref{tab:MatResults}. For both analyses, the fiber geometry equals the mean value of the fiber length and fiber diameter. The fiber mass fraction is set to $30 \%$ and the fiber orientation is $0^\circ$. The results of the numerical analysis are provided in Table \ref{tab:Comp}. As the coefficients $C_{12}$ and $C_{21}$ can be calculated individually in this case, the given value of $C_{12}$ represents the mean of $C_{12}$ and $C_{21}$. The comparison of the numerical and analytical results agrees very well for the coefficients $C_{12}$, $C_{22}$, and $C_{66}$. This is independent of the boundary condition as well as the window size. In contrast to this, the coefficient $C_{11}$ is approximated higher by using the analytical models of Tandon and Weng as well as Halpin and Tsai than the numerical simulations indicate. Finally, the mean values of $C_{16}, C_{26}, C_{61}$, and $C_{62}$ based on the numerical simulations are close to zero. This fits the analytical results for fibers aligned with the symmetry axis. Therefore, the analytical and numerical results show an overall good agreement. 

\begin{table}[tb]
    \centering
    \begin{tabular}{lllllll} 
    \toprule
    & BC & $\overline{C_{11}} \ [\SI{}{\giga\pascal}]$& $\overline{C_{12}} \ [\SI{}{\giga\pascal}]$ & $\overline{C_{22}} \ [\SI{}{\giga\pascal}]$& $\overline{C_{66}} \ [\SI{}{\giga\pascal}]$  \\  
    \midrule 
    Tandon-Weng & & $13.0$ & $1.59$ & $4.18$ &  $1.31$\\
    Halpin-Tsai & & $12.0$ & $1.63$ & $4.34$ &  $1.30$\\ 
    \midrule 
    \SI{250}{\micro\metre} & KUBC & $11.8$ & $1.56$ & $4.36$  & $1.35$\\
    \SI{250}{\micro\metre} & SUBC & $6.43$ & $1.48$ & $3.91$  & $1.16$\\
    \midrule
    \SI{500}{\micro\metre} & KUBC & $10.2$ & $1.57$ & $4.14$  & $1.27$\\
    \SI{500}{\micro\metre} & SUBC & $7.11$ & $1.53$ & $3.92$ & $1.18$\\
    \midrule 
    \SI{750}{\micro\metre} & KUBC & $9.80$ & $1.56$ & $4.08$  & $1.24$\\
    \SI{750}{\micro\metre} & SUBC & $7.59$ & $1.59$ & $3.93$ &  $1.18$\\
    \bottomrule
    \end{tabular}
    \caption{Comparison of the mean values for the engineering constants based on the analytical and numerical analysis.}
    \label{tab:Comp}
\end{table}

\subsection{Correlation analysis}
\label{sec:Corr}

\subsubsection{Moving window}

\begin{figure}[tb]
\begin{subfigure}[t]{0.30\textwidth}
\includegraphics[width=\textwidth]{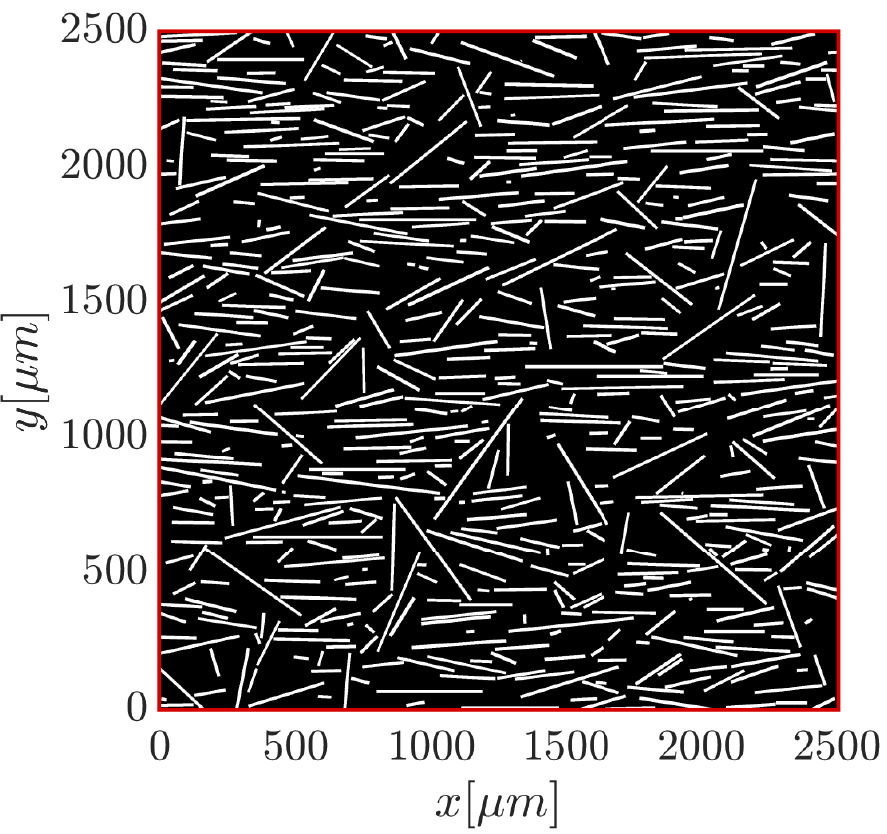}
\subcaption{\raggedright Created microstructure with a size of $\SI{2500}{\micro\metre} \times \SI{2500}{\micro\metre}$.}
\label{fig:Mic}
\end{subfigure}
\begin{subfigure}[t]{0.30\textwidth}
\includegraphics[width=\textwidth]{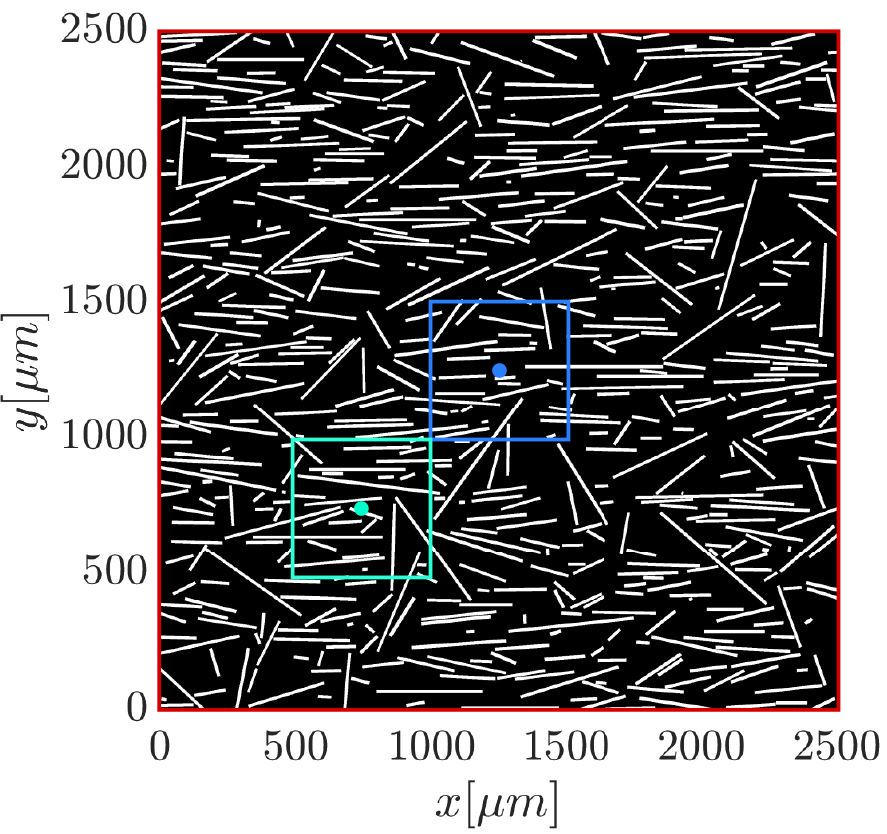}
\subcaption{\raggedright Extracting windows of $\SI{500}{\micro\metre} \times \SI{500}{\micro\metre}$ from the original microstructure at different locations.}
\label{fig:MovingWindow}
\end{subfigure}
\begin{subfigure}[t]{0.30\textwidth}
\includegraphics[width=\textwidth]{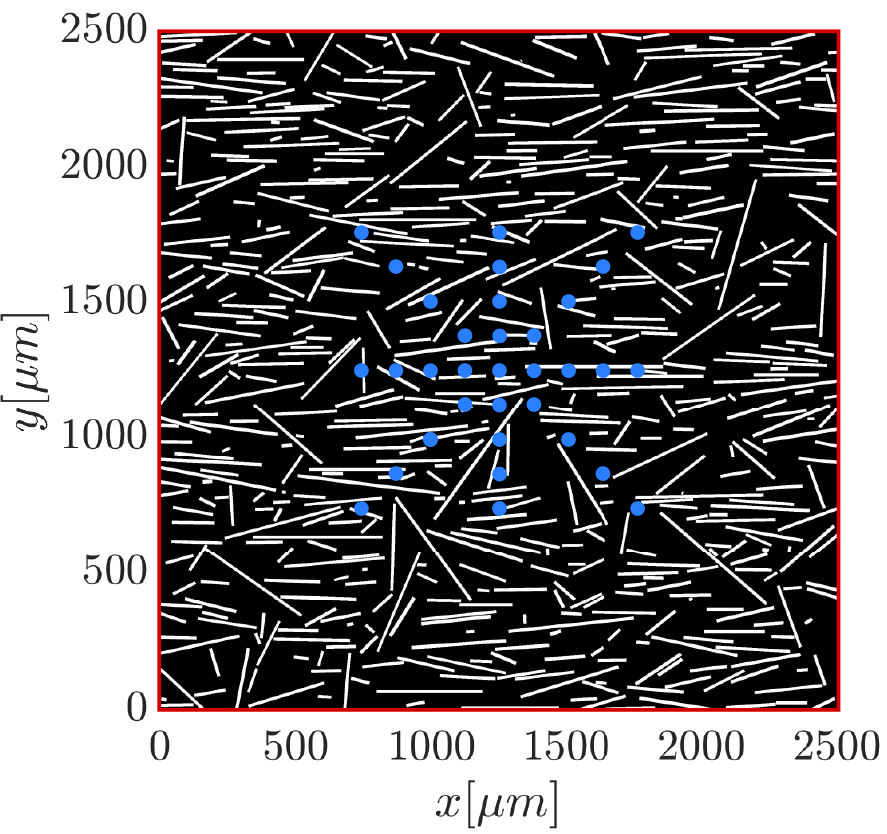}
\subcaption{\raggedright Location of the center points of each window that is used to analyze the correlation of the elasticity coefficients.}
\label{fig:Corr}
\end{subfigure}
\caption{Moving window procedure}
\label{fig:MovingWindowall}
\end{figure} 

The moving window method is used to characterize the random local properties of composites \cite{Sena.2013,Baxter.2000,GrahamBrady.2003}. In this context, a window of a microstructure is moved to different locations and marks a rectangular part of the microstructure. The procedure is depicted in Figure \ref{fig:MovingWindow} for a microstructure with a size of $\SI{2500}{\micro\metre} \times \SI{2500}{\micro\metre}$ and window sizes of $\SI{500}{\micro\metre} \times \SI{500}{\micro\metre}$. With respect to the correlation analysis, this technique can be used to analyze the local dependence of the material properties by evaluating overlapping windows. If the windows do not overlap the correlation should be zero, as their microstructures are independent of each other.  

In this study the basis of the correlation analysis is a microstructure of $\SI{2500}{\micro\metre} \times \SI{2500}{\micro\metre}$. To analyze the scale dependence of the correlation structure the window size is varied. Here, the maximum possible window size regarding a microstructure edge length of \SI{2500}{\micro\metre} is \SI{833}{\micro\metre}. In this case, it is possible to place three windows next to each other without any overlap. With a maximum window size of \SI{833}{\micro\metre} and in accordance with the analysis presented so far the correlation analysis is done using again window sizes of \SI{250}{\micro\metre}, \SI{500}{\micro\metre}, and \SI{750}{\micro\metre}. With respect to an average fiber length of \SI{260}{\micro\metre} this domain can be assigned to the mesoscale. This is also indicated by the results of the elasticity tensor properties presented in Section \ref{sec:Overall} which clearly show a scale dependence of the elasticity coefficients for window sizes up to \SI{750}{\micro\metre}.

To determine the correlation structure the dimensionless correlation parameter is calculated by evaluating Eq. (\ref{eqn:DimCovDis}) based on 33 extracted windows at different locations from the same microstructure, see Figure \ref{fig:Corr}. Starting from the center point of the microstructure the window is moved to the left and right as well as to the top and bottom four times. The distance between the equidistantly arranged center points of the extracted windows in the same direction is a quarter of the current window size. Therefore, the outer windows and the window in the center of the microstructure do not overlap. In the same way, windows are placed along the diagonals of the microstructure. Here, the distance between two center points is $\sqrt{2} l_{\text{window}}$. This procedure is repeated for a total of 500 microstructures to ensure convergence of the dimensionless correlation parameters \cite{Sena.2013,Soize.2008}. 

\subsubsection{Cross-correlation}

\begin{figure}[t]
\centering
\includegraphics[width=0.92\textwidth]{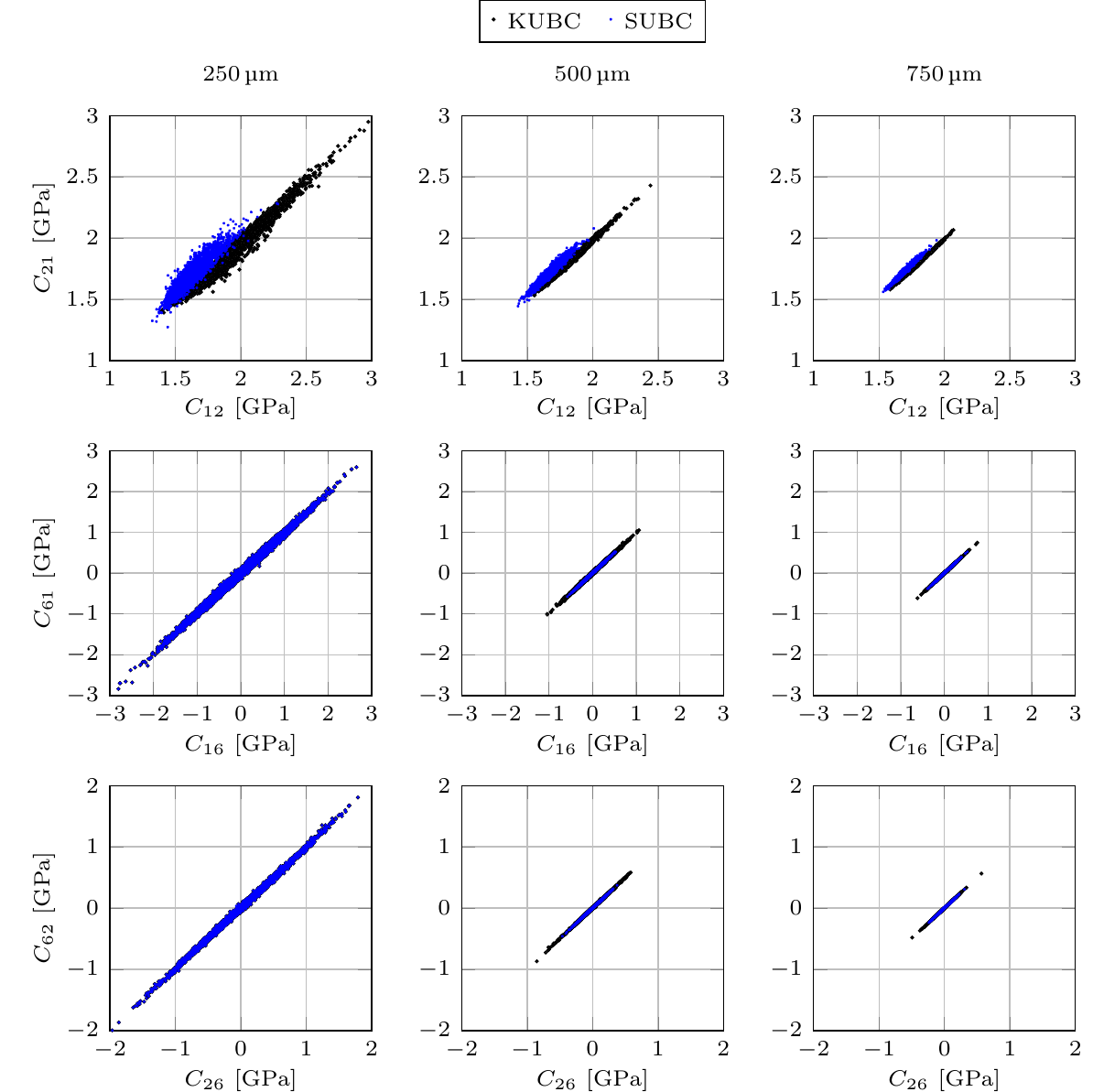}
\caption{Symmetry analysis of the elasticity tensor.}
\label{fig:sym}
\end{figure}

In this section, the correlation structure of the elasticity tensor is analyzed in detail. The main focus lays on the cross-correlation of the elasticity coefficients with respect to the boundary condition and the window size. First, the symmetry of the elasticity tensor is analyzed. Of particular interest are the elements $C_{12}$ and $C_{21}$. Furthermore, as the varying fiber orientation leads to the anisotropy of the material the elements $C_{16}$ and $C_{61}$ as well as $C_{26}$ and $C_{62}$ are taken into account as well. Therefore, Figure \ref{fig:sym} shows the results of each simulation, where $C_{12}$ is plotted against $C_{21}$, $C_{16}$ is plotted against $C_{61}$, and $C_{26}$ is plotted against $C_{62}$, respectively. The correlation between these elasticity coefficients is clearly indicated by a strong alignment of the points. Furthermore, the values of $C_{16}$, $C_{26}$, $C_{61}$, and $C_{62}$ fluctuate around zeros, which leads to the conclusion that the mean value of these elements is close to zero. This behavior is independent of the boundary condition. However, the anisotropic effect ($C_{16}, C_{26}, C_{61},C_{62} \neq 0$) starts to vanish with an increasing window size. This meets the assumption of overall transversely-isotropic material properties.

\begin{figure}
    \centering
    \includegraphics[width=0.9\textwidth]{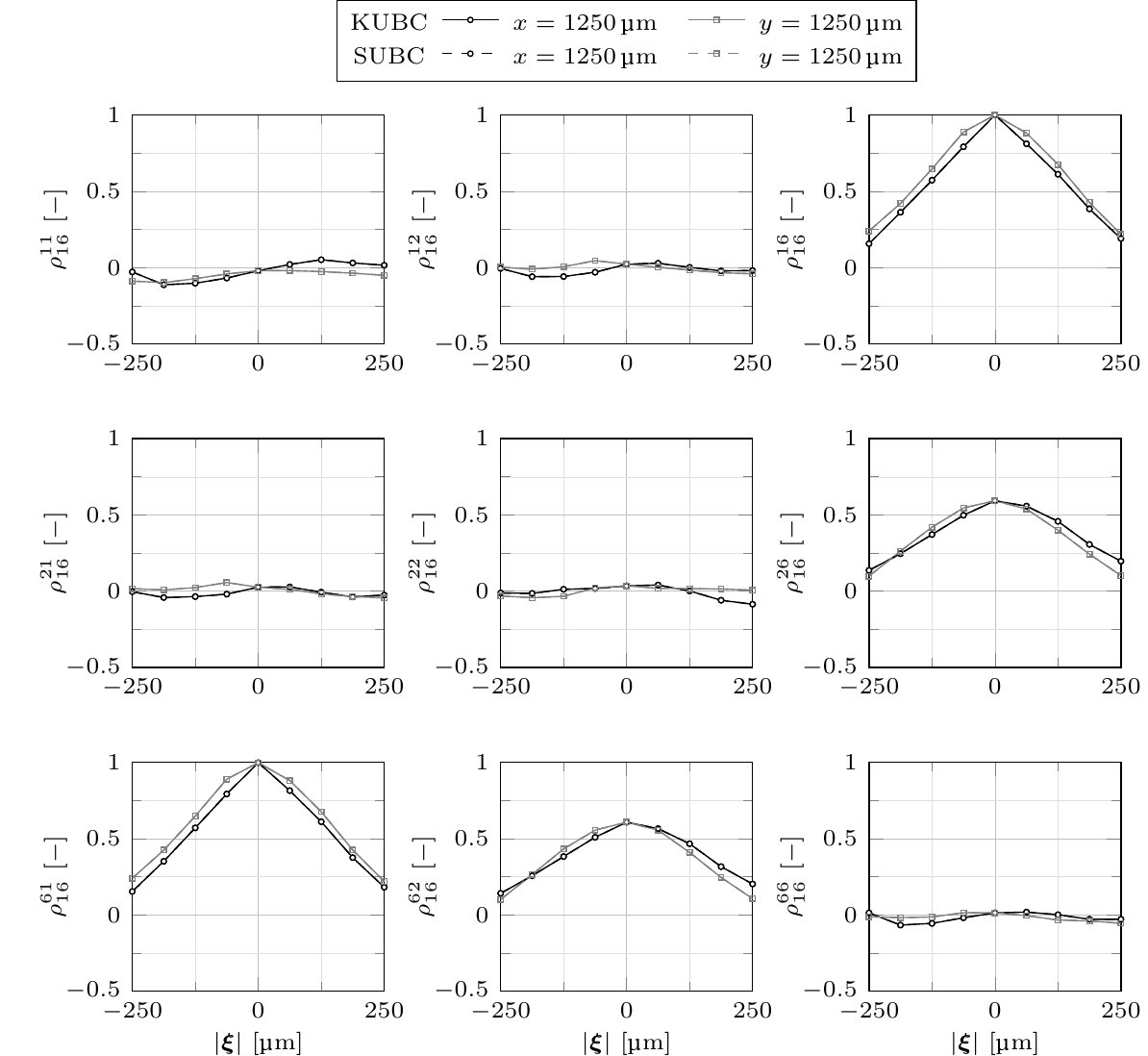}
    \caption{Correlation analysis of the elasticity tensor based on the element $C_{16}$ for a window size of \SI{250}{\micro\metre}.}
    \label{fig:C13}
\end{figure}

For analyzing the correlation structure of the anisotropic effect more in detail, Figure \ref{fig:C13} depicts the correlation structure of $C_{16}$ with respect to a window size of \SI{250}{\micro\metre}. Here, the lower index gives the reference for the dimensionless correlation parameter and the upper index refers to the elasticity coefficient that is calculated based on the moving window. Therefore, for $\rho_{16}^{11}$ $C_{16}$ is calculated for the window extracted of the microstructure center and $C_{11}$ is determined for the extract of the moving window. Furthermore, $\vert\boldsymbol{\xi}\vert$ gives the distance between the moving window and the center point of the microstructre.

\begin{figure}[t]
\centering
\includegraphics[width=0.92\textwidth]{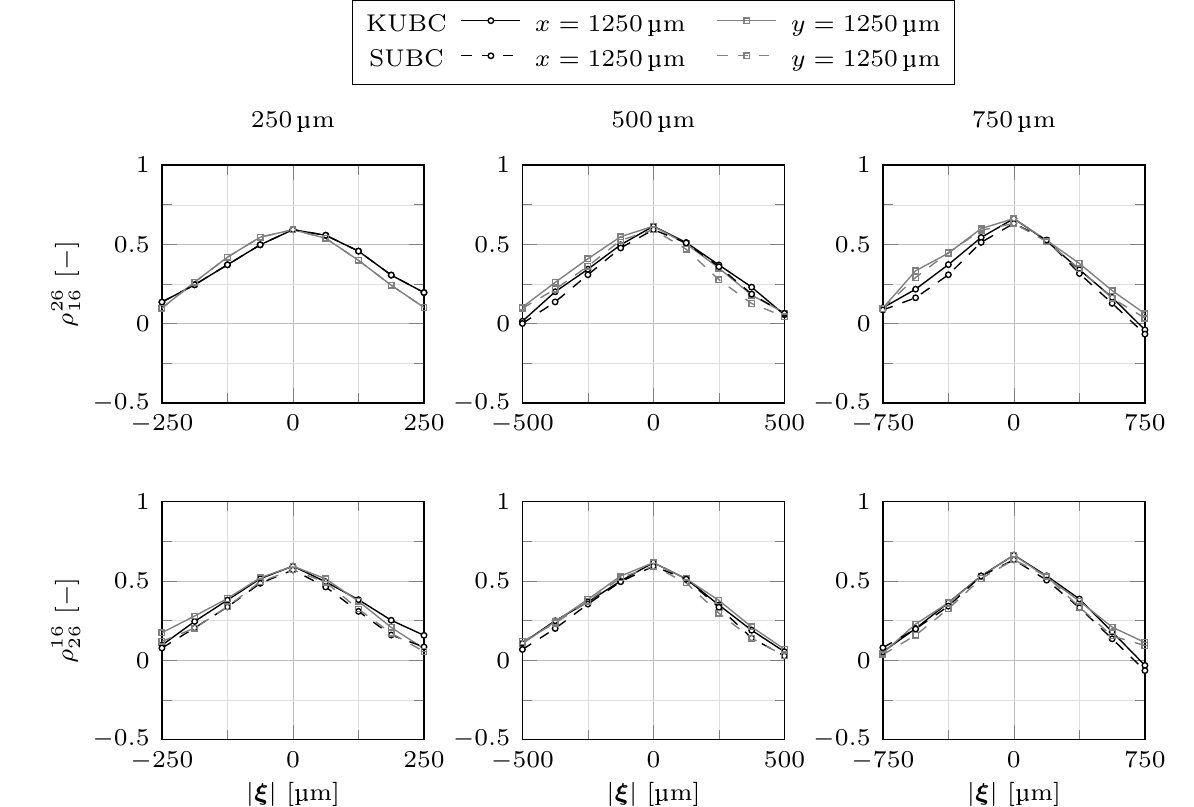}
\caption{Comparison of $\rho_{16}^{26}$ and $\rho_{26}^{16}$.}
\label{fig:Symmetry2}
\end{figure}

The following essential characteristics can be derived. First, the correlation structure with respect to $C_{16}$ is independent of the boundary condition. Second, the symmetry properties of the elasticity tensor can also be found within the correlation structure as  $\rho_{16}^{16}$ is almost identical to $\rho_{16}^{61}$. The same holds for $\rho_{16}^{26}$ and $\rho_{16}^{62}$. Finally, there are only cross-correlations between the elements $C_{16}$ and $C_{61}$ due to the symmetry as well as between $C_{16}$ and $C_{26}$ and $C_{16}$ and $C_{62}$, respectively. All other elements are uncorrelated to $C_{16}$. As indicated by Figure \ref{fig:Symmetry2}, that shows the dimensionless correlation parameter between $C_{16}$ and $C_{26}$ for all three window sizes as well as both boundary conditions, these conclusions are also independent of the window size. Furthermore, the dimensionless correlation parameter is calculated in two different ways. First, the reference is $C_{16}$ and $C_{26}$ is calculated for the moving window. Subsequently, the values are exchanged, which means that the reference is $C_{26}$ and $C_{16}$ is calculated for the moving window. Based on the identical correlations for both variants,
\begin{equation}
\rho_{16}^{26} = \rho_{26}^{16}
\end{equation}
holds under the condition, that a sufficient number of realizations is taken as a basis. This leads to the conclusion that the correlation structure shows the same symmetry properties as the elasticity tensor and therefore,
\begin{equation}
\rho_a^b = \rho_b^a
\end{equation}
can be assumed. Hence, the remaining number of independent correlation parameters can be reduced to 
\begin{equation}
\rho_{11}^{12}, \ \rho_{11}^{22}, \ \rho_{11}^{66}, \ \rho_{12}^{22}, \ \rho_{12}^{66}, \ \rho_{22}^{66}.
\end{equation}
These conclusions are also valid for the correlation structure based on a plane strain state as shown in Figures \ref{fig:comp1} and \ref{fig:comp3}.

\begin{figure}[t]
\centering
\includegraphics[width=0.92\textwidth]{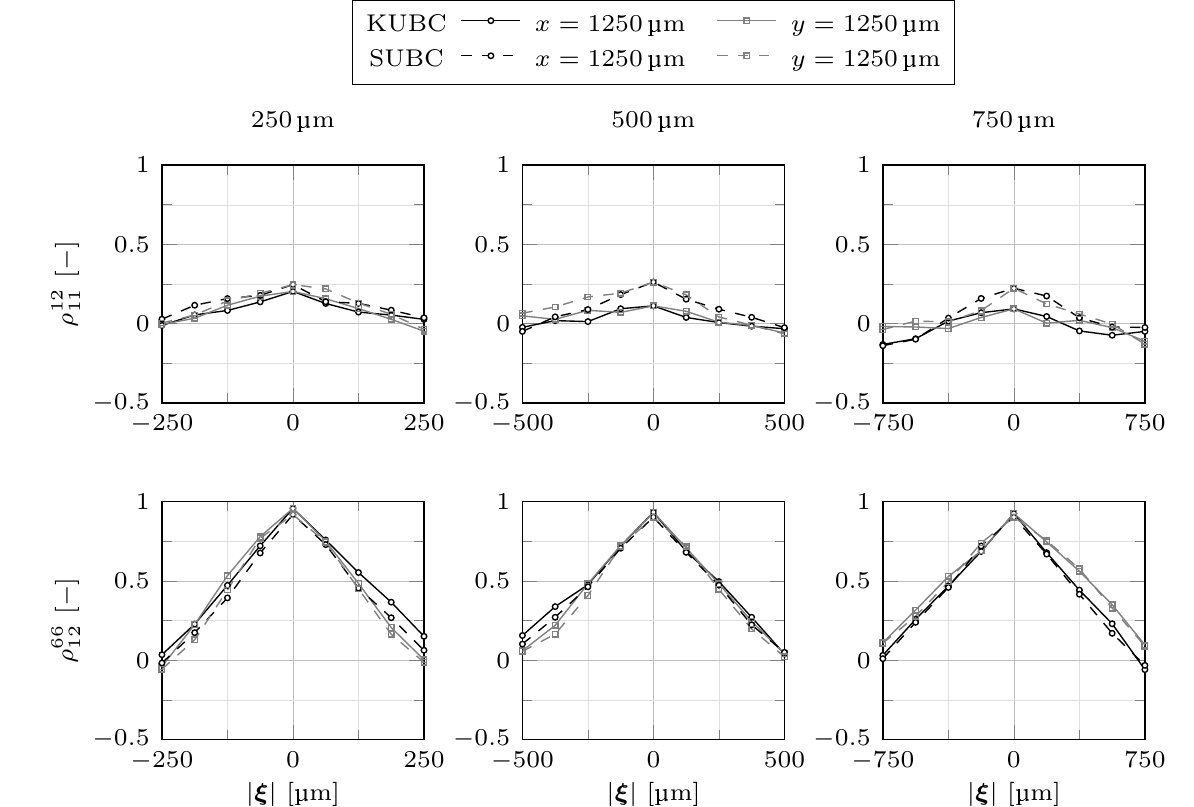}
\caption{Dimensionless correlation parameters that are independent of the window size.}
\label{fig:C11_cross}
\end{figure}

\begin{figure}[p]
\centering
\includegraphics[width=0.92\textwidth]{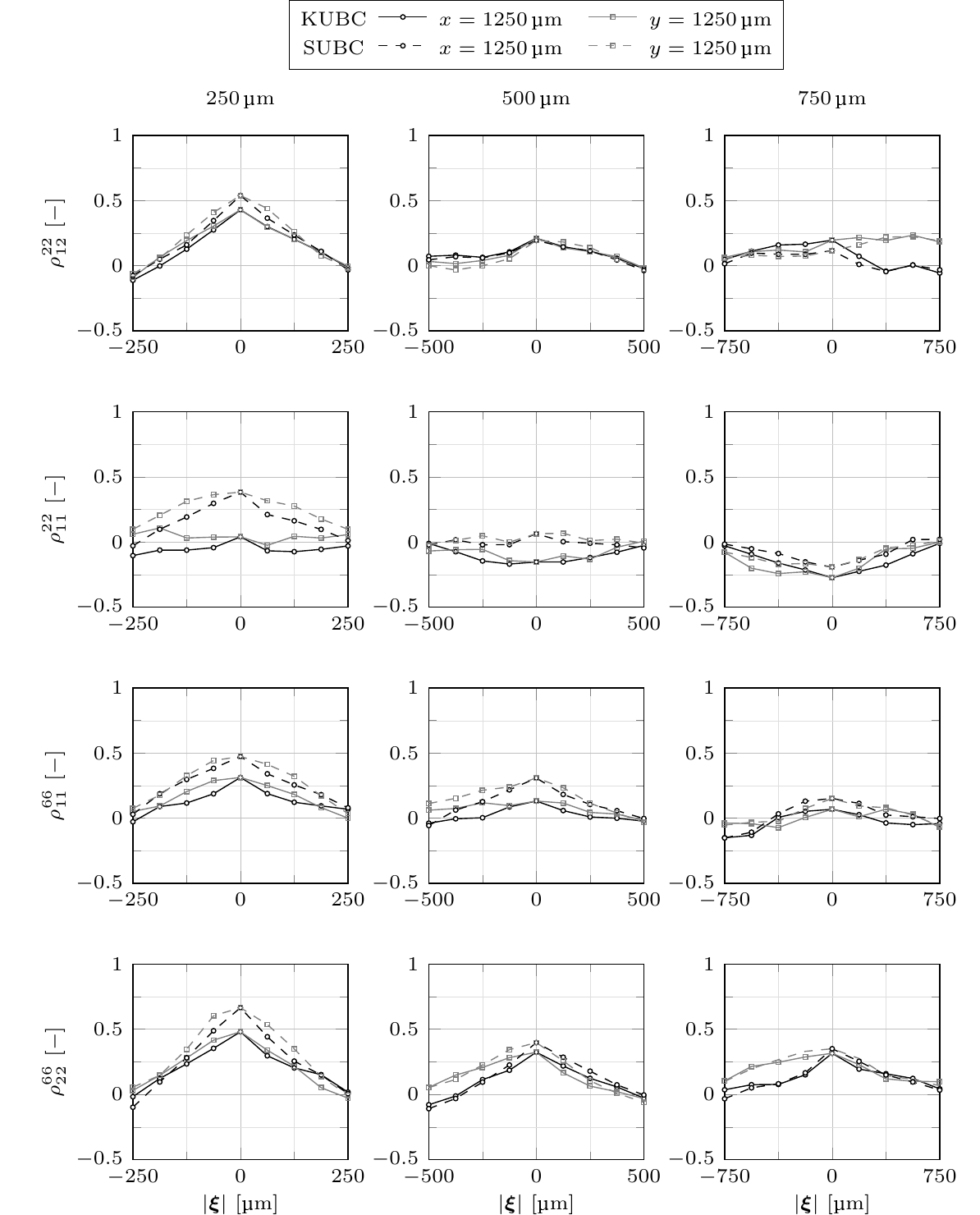}
\caption{Dimensionless correlation parameters that depend on the window size.}
\label{fig:C12_cross}
\end{figure}

The results of the cross-correlation for the remaining elasticity coefficients are presented in Figures \ref{fig:C11_cross} and \ref{fig:C12_cross}. Here, the dimensionless correlation parameter is again divided into two groups. For one the correlation is independent of the window size (Figure \ref{fig:C11_cross}). The other group shows a decreasing correlation for an increasing window size, which is indicated by a decreasing value for $\vert \boldsymbol{\xi}\vert = 0$. The overall shape of the correlation is not necessarily affected. One example is $\rho_{11}^{66}$. If the correlation depends on the window size, it is only relevant as long as the structure is represented by a SVE. In the case of a RVE the correlation vanishes.

These results also meet the theoretical framework as there is a clear connection between $C_{11}$ and $C_{12}$ via the Poisson ration $\nu_{12}$. However, the two elasticity coefficient $C_{11}$ and $C_{66}$ are independent on the macroscale.

The same correlation structure can be also derived on the basis of a plane strain assumption, see Figures \ref{fig:comp2} and \ref{fig:comp4}. However, the correlation functions can not be approximated by the same function. Therefore, when using second-order random fields for numerical simulations of components the correlation functions must be determined with respect to the overall load assumption.

\section{Conclusion}
In the presented work first, an analytical analysis is performed. Based on the material models by Tandon and Weng as well as Halpin and Tsai the influence of the different geometrical properties like fiber length, fiber diameter, fiber orientation, and fiber volume fraction on the material properties is investigated. The results show that the Young's Modulus $E_1$ is the most affected material property by the geometrical properties. This is supported by the material model of Halpin and Tsai, where only $E_1$ is modeled as a function of the fiber length and fiber diameter. The main influencing parameters of $E_1$ are the fiber length, fiber orientation, and fiber volume fraction. The remaining engineering constants are only slightly affected by the geometrical fiber properties. However, the fiber orientation, as well as the fiber volume fraction, also have a significant effect on these constants. This can also be found when analyzing the elasticity coefficients. Furthermore, the distribution of the resulting elasticity tensor, as well as the elasticity coefficients, can be approximated by the same function as the probability density function of the influencing parameter itself.

In the next step, this analysis is repeated on a numerical basis. However, there is one important difference between the numerical and analytical analyses. The numerical model consists of several fibers of different length, diameter, and orientation. In contrast to that the analytical analysis is based on the assumption that all fibers have the same length, diameter, and orientation. Furthermore, by evaluating the numerical model each element of the elasticity tensor can be calculated individually. Therefore, the symmetry of this tensor is not presupposed, but the numerical results clearly confirm the symmetry of the resulting elasticity tensor. Furthermore, the overall transversely-isotropic material properties are affirmed by the numerical simulations as the mean of the elasticity coefficients $C_{16}$ and $C_{26}$ equals zero and the standard deviation decreases with increasing window size.

Summarizing the numerical simulations show a good agreement with the results of the analytical investigation. The material properties of the microstructure is influenced significantly by the fiber orientation and the fiber volume fraction. Furthermore, on the mesoscale, the distribution of the elasticity coefficients can be approximated with the same probability distribution as the influencing parameter itself. However, with increasing window size, the distribution starts to shift to a normal distribution. 

Regarding the numerical correlation analysis, the symmetry of elasticity tensor can also be found within the correlation structure of the elasticity coefficients. In addition, the dimensionless correlation parameters can be divided into two independent groups. The first group gives the correlation between the non zero elasticity coefficients of transversely-isotropic material behavior. The second group comprises only a correlation of $C_{16}$ and $C_{26}$. 

It can be concluded, that the correlation structure of the elasticity tensor for SFRC can be approximated by numerical simulations on the mesoscale and the results are independent of the chosen boundary conditions. However, the results depend not only on the window size but also on the overall assumption of a plane stress and a plane strain state in case of a two-dimensional model, respectively.

\appendix

\section{Analytical results}
\label{sec:Results_ana}

Below the remaining results of the analytical analysis are shown. This includes the influence of the fiber diameter and fiber orientation on the engineering constants as well as the elasticity coefficients based on the material model by Tandon and Weng. Furthermore, the results based on the material model by Halpin and Tsai are given. This covers the Young's Modulus $E_1$ as well as the elasticity coefficients, both in dependence of the fiber length, diameter and fiber orientation.

\pgfplotsset{scaled y ticks=false}

\begin{figure}[p]
\centering
\includegraphics[width=0.92\textwidth]{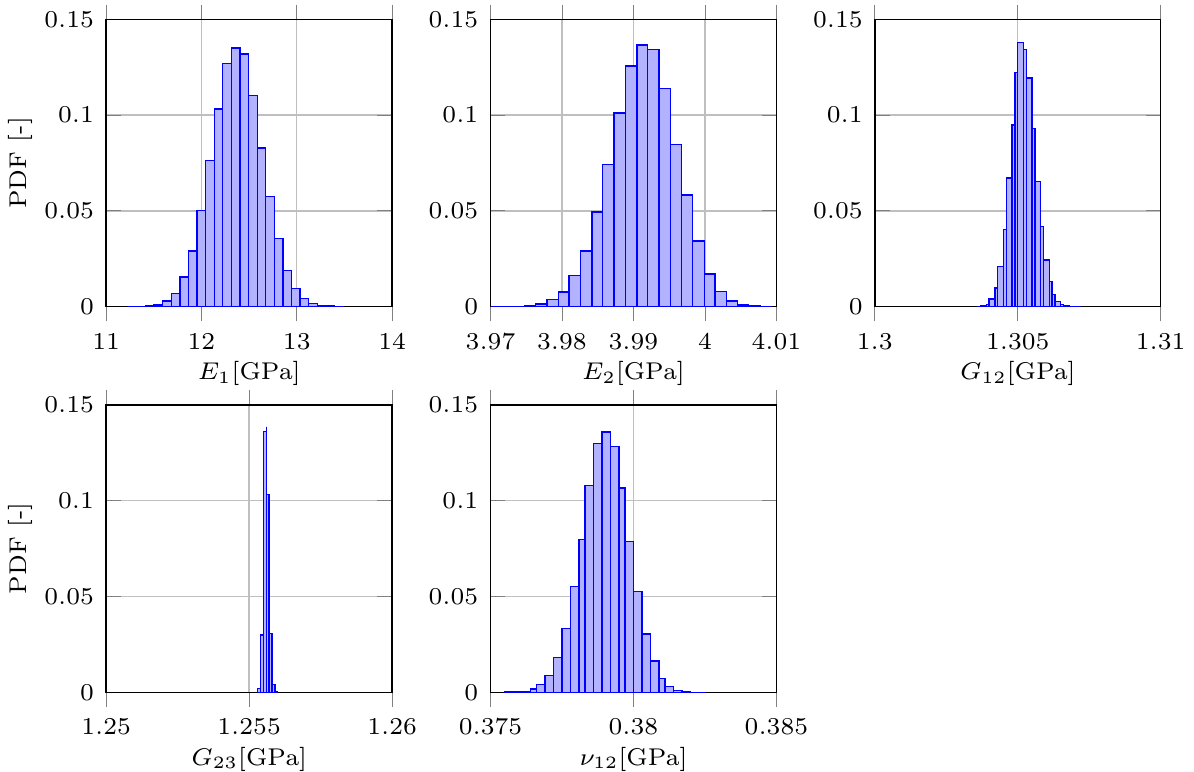}
\caption{Engineering constants due to a varying fiber diameter calculated with the Tandon-Weng material model.}
\end{figure}

\begin{figure}[p]
\centering
\includegraphics[width=0.92\textwidth]{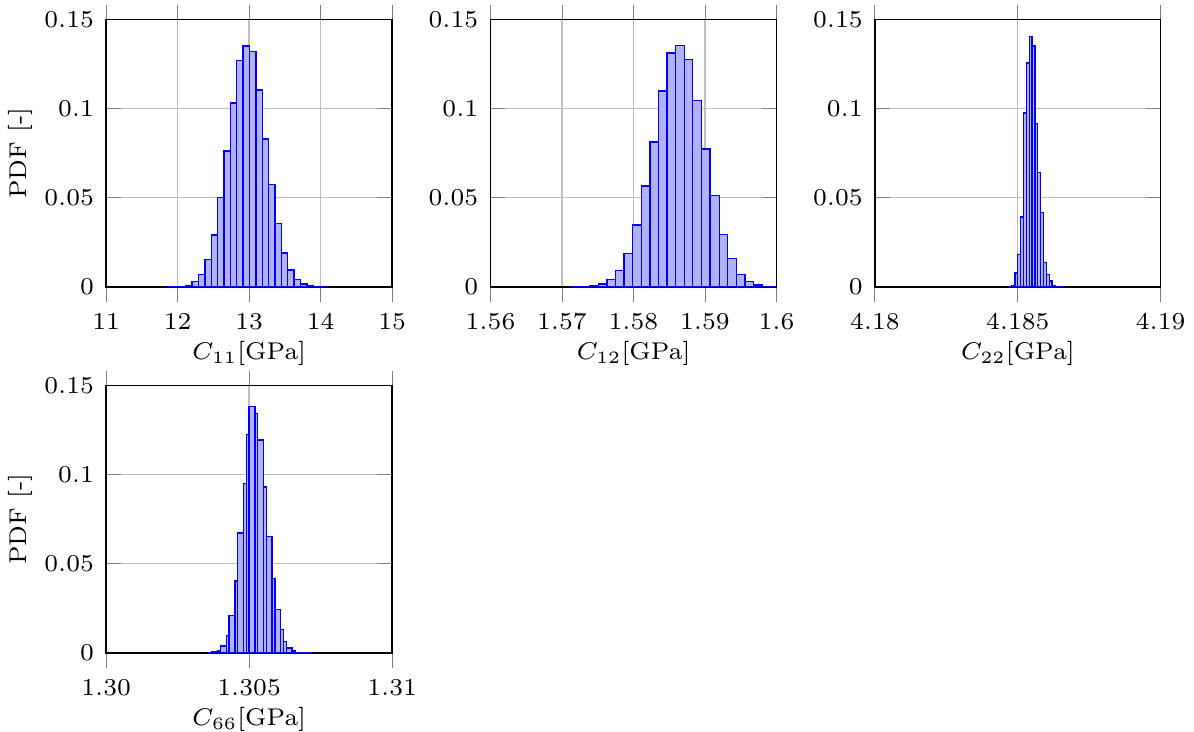}
\caption{Elasticity coefficients due to a varying fiber diameter calculated with the Tandon-Weng material model.}
\end{figure}

\begin{figure}[p]
\centering
\includegraphics[width=0.92\textwidth]{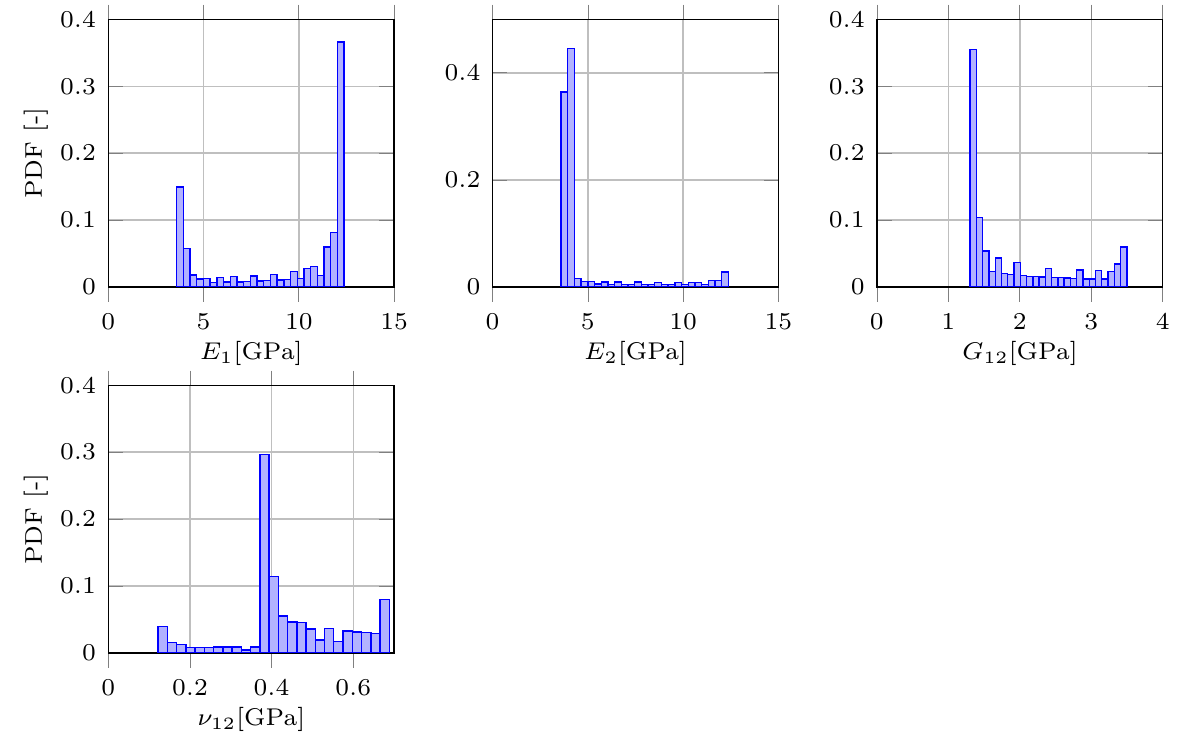}
\caption{Engineering constants due to a varying fiber orientation calculated with the Tandon-Weng material model.}
\end{figure}

\begin{figure}[p]
\centering
\includegraphics[width=0.92\textwidth]{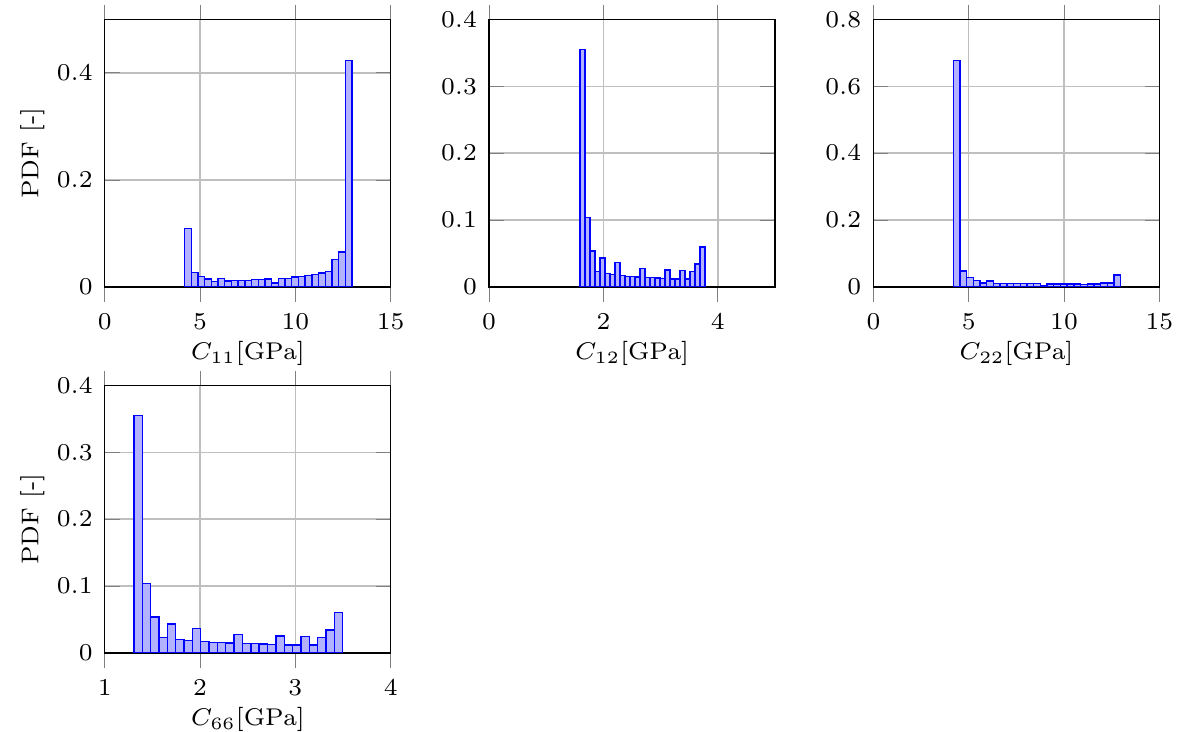}
\caption{Elasticity coefficients due to a varying fiber orientation calculated with the Tandon-Weng material model.}
\end{figure}

\begin{figure}[p]
\centering
\includegraphics[width=0.92\textwidth]{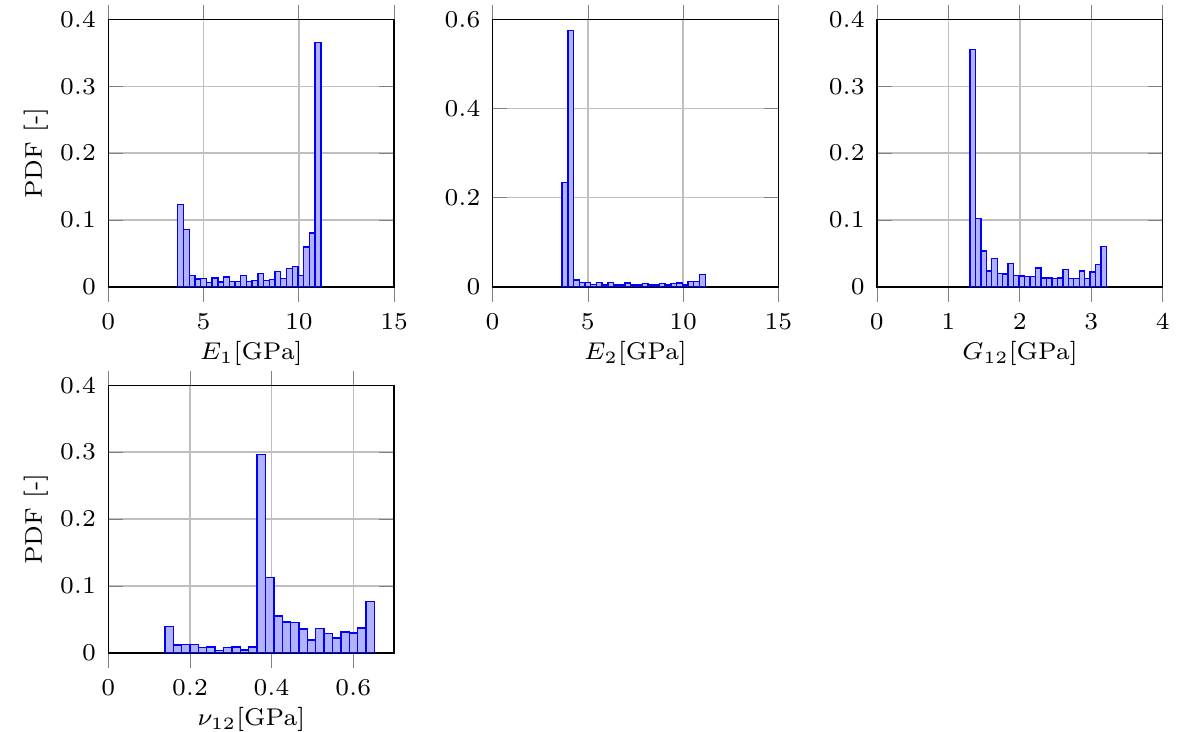}
\caption{Engineering constants due to a varying fiber orientation calculated with the Halpin-Tsai material model.}
\end{figure}

\begin{figure}[p]
\centering
\includegraphics[width=0.92\textwidth]{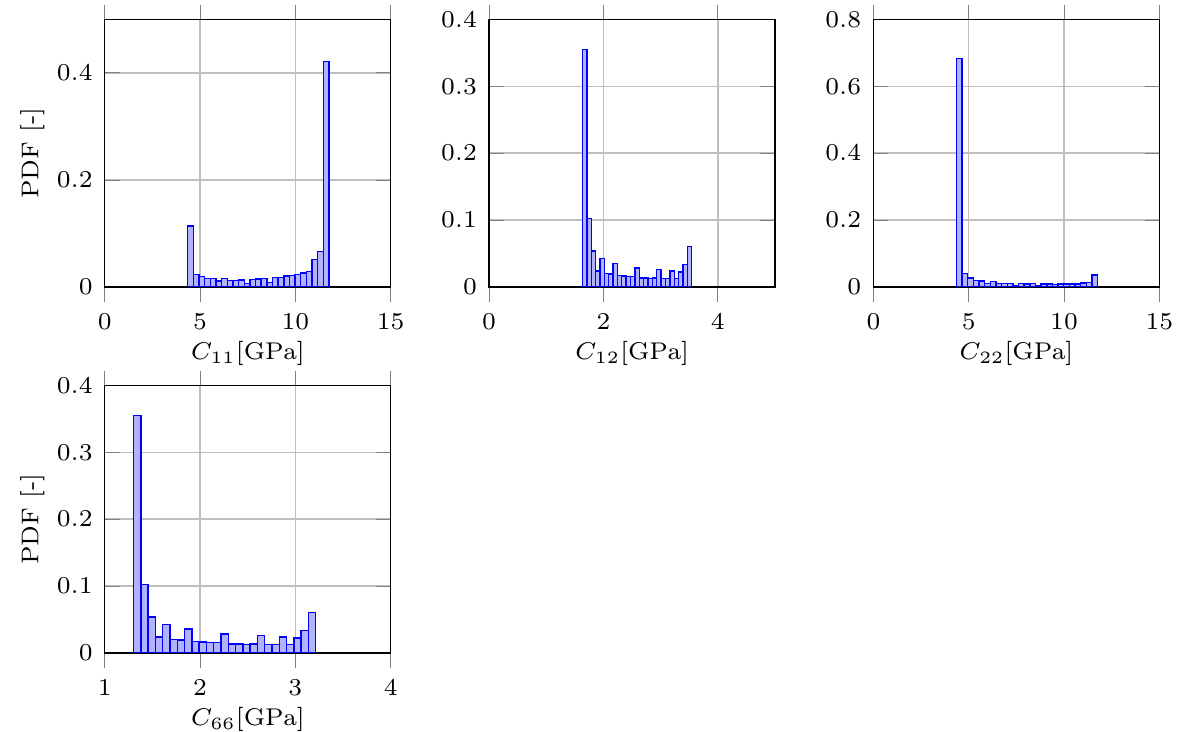}
\caption{Elasticity coefficients due to a varying fiber orientation calculated with the Halpin-Tsai material model.}
\end{figure}

\clearpage

\section{Numerical analysis}
\subsection{Determination of the elasticity tensor elements}
\label{sec:Calc_E}
To calculate the elasticity coefficients a system of equations can be formulated based on the three independent load cases. Hooke's law for a finite volume element reads
\begin{equation}
    \langle \boldsymbol{\sigma}\rangle = \mathbb{C}^{\text{eff}} : \langle \boldsymbol{\epsilon} \rangle.
    \tag{\ref{eqn:Ceff}}
\end{equation}
Reducing this formulation to the 2D case and using Voigt notation leads to
\begin{equation}
    \begin{bmatrix}
        \langle \sigma_1 \rangle \\
        \langle \sigma_2 \rangle \\
        \langle \sigma_6 \rangle 
    \end{bmatrix}
    =
    \begin{bmatrix}
        C_{11}^{\text{eff}} & C_{12}^{\text{eff}} & C_{16}^{\text{eff}} \\
        C_{21}^{\text{eff}} & C_{22}^{\text{eff}} & C_{26}^{\text{eff}} \\
        C_{61}^{\text{eff}} & C_{62}^{\text{eff}} & C_{66}^{\text{eff}} 
    \end{bmatrix}
    \begin{bmatrix}
        \langle \epsilon_1 \rangle \\
        \langle \epsilon_2 \rangle \\
       2 \langle \epsilon_6 \rangle 
    \end{bmatrix}.
\end{equation}
Each stress component can be written in an individual equation
\begin{equation}
    \begin{bmatrix}
        \langle \sigma_1 \rangle
    \end{bmatrix}
    =
    \begin{bmatrix}
        C_{11}^{\text{eff}} & C_{12}^{\text{eff}} & C_{16}^{\text{eff}}
    \end{bmatrix}
    \begin{bmatrix}
        \langle \epsilon_1 \rangle \\
        \langle \epsilon_2 \rangle \\
       2 \langle \epsilon_6 \rangle 
    \end{bmatrix}
\end{equation}
which can be rearranged to
\begin{equation}
    \begin{bmatrix}
        \langle \sigma_1 \rangle
    \end{bmatrix}
    =
    \begin{bmatrix}
        \langle \epsilon_1 \rangle &
        \langle \epsilon_2 \rangle &
       2 \langle \epsilon_6 \rangle 
    \end{bmatrix}
        \begin{bmatrix}
        C_{11}^{\text{eff}} \\
        C_{12}^{\text{eff}} \\
        C_{16}^{\text{eff}}
    \end{bmatrix}.
\end{equation}
Now the individual load cases can be summarized in this system of equations
\begin{equation}
    \begin{bmatrix}
        \langle \sigma_1^{\text{LC1}} \rangle \\
        \langle \sigma_1^{\text{LC2}} \rangle \\
        \langle \sigma_1^{\text{LC3}} \rangle
    \end{bmatrix}
    =
    \begin{bmatrix}
        \langle \epsilon_1^{\text{LC1}} \rangle & \langle \epsilon_2^{\text{LC1}} \rangle & 2 \langle \epsilon_6^{\text{LC1}} \rangle \\
        \langle \epsilon_1^{\text{LC2}} \rangle & \langle \epsilon_2^{\text{LC2}} \rangle & 2 \langle \epsilon_6^{\text{LC2}} \rangle \\
        \langle \epsilon_1^{\text{LC3}} \rangle & \langle \epsilon_2^{\text{LC3}} \rangle & 2 \langle \epsilon_6^{\text{LC3}} \rangle 
    \end{bmatrix}
        \begin{bmatrix}
        C_{11}^{\text{eff}} \\
        C_{12}^{\text{eff}} \\
        C_{16}^{\text{eff}}
    \end{bmatrix}
\end{equation}
which can be solved by
\begin{equation}
        \begin{bmatrix}
        C_{11}^{\text{eff}} \\
        C_{12}^{\text{eff}} \\
        C_{16}^{\text{eff}}
    \end{bmatrix}
    =
    \begin{bmatrix}
        \langle \epsilon_1^{\text{LC1}} \rangle & \langle \epsilon_2^{\text{LC1}} \rangle & 2 \langle \epsilon_6^{\text{LC1}} \rangle \\
        \langle \epsilon_1^{\text{LC2}} \rangle & \langle \epsilon_2^{\text{LC2}} \rangle & 2 \langle \epsilon_6^{\text{LC2}} \rangle \\
        \langle \epsilon_1^{\text{LC3}} \rangle & \langle \epsilon_2^{\text{LC3}} \rangle & 2 \langle \epsilon_6^{\text{LC3}} \rangle 
    \end{bmatrix}^{-1}
    \begin{bmatrix}
        \langle \sigma_1^{\text{LC1}} \rangle \\
        \langle \sigma_1^{\text{LC2}} \rangle \\
        \langle \sigma_1^{\text{LC3}} \rangle
    \end{bmatrix}.
\end{equation}
Expanding this procedure to the remaining stress components $\langle \sigma_2 \rangle$ and $\langle \sigma_6 \rangle$ the elasticity coefficients can be derived from
\begin{equation}
        \begin{bmatrix}
        C_{n1}^{\text{eff}} \\
        C_{n2}^{\text{eff}} \\
        C_{n6}^{\text{eff}}
    \end{bmatrix}
    =
    \begin{bmatrix}
        \langle \epsilon_1^{\text{LC1}} \rangle & \langle \epsilon_2^{\text{LC1}} \rangle & 2 \langle \epsilon_6^{\text{LC1}} \rangle \\
        \langle \epsilon_1^{\text{LC2}} \rangle & \langle \epsilon_2^{\text{LC2}} \rangle & 2 \langle \epsilon_6^{\text{LC2}} \rangle \\
        \langle \epsilon_1^{\text{LC3}} \rangle & \langle \epsilon_2^{\text{LC3}} \rangle & 2 \langle \epsilon_6^{\text{LC3}} \rangle 
    \end{bmatrix}^{-1}
    \begin{bmatrix}
        \langle \sigma_n^{\text{LC1}} \rangle \\
        \langle \sigma_n^{\text{LC2}} \rangle \\
        \langle \sigma_n^{\text{LC3}} \rangle
    \end{bmatrix}
\end{equation}
with $n=1,2,6$.
\newpage
\subsection{Results based on a plane strain assumption}
\label{sec:strain}

The following figures show the results for the correlation analysis assuming a plane strain state. 

\begin{figure}[h]
\centering
\includegraphics[width=0.92\textwidth]{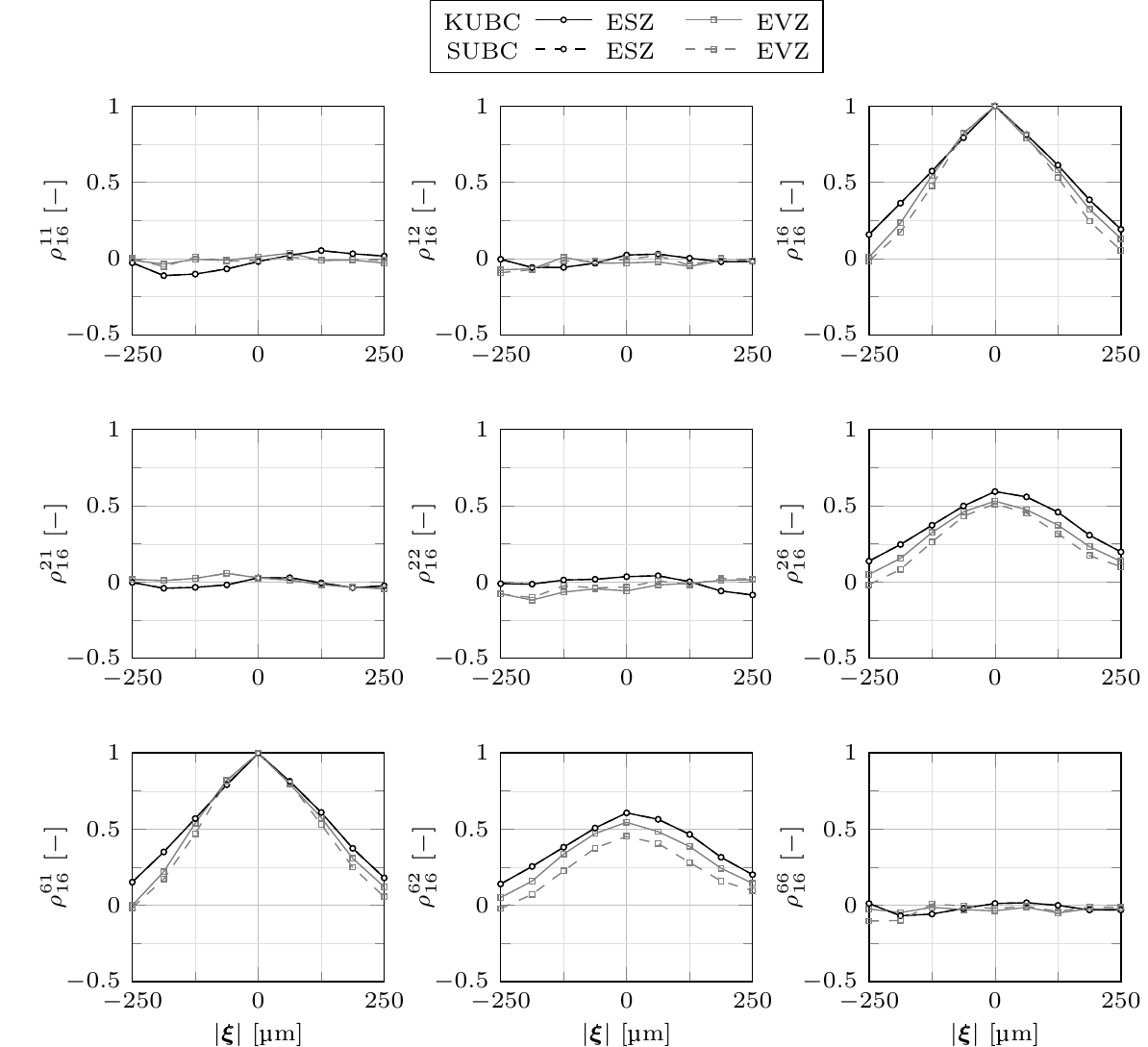}
\caption{Comparison of the dimensionless correlation parameters based on $C_{16}$ assuming plane strain and plane stress states for a window size of \SI{250}{\micro \metre} at $x = \SI{1250}{\micro\metre}$.}
\label{fig:comp1}
\end{figure}

\begin{figure}
\centering
\includegraphics[width=0.92\textwidth]{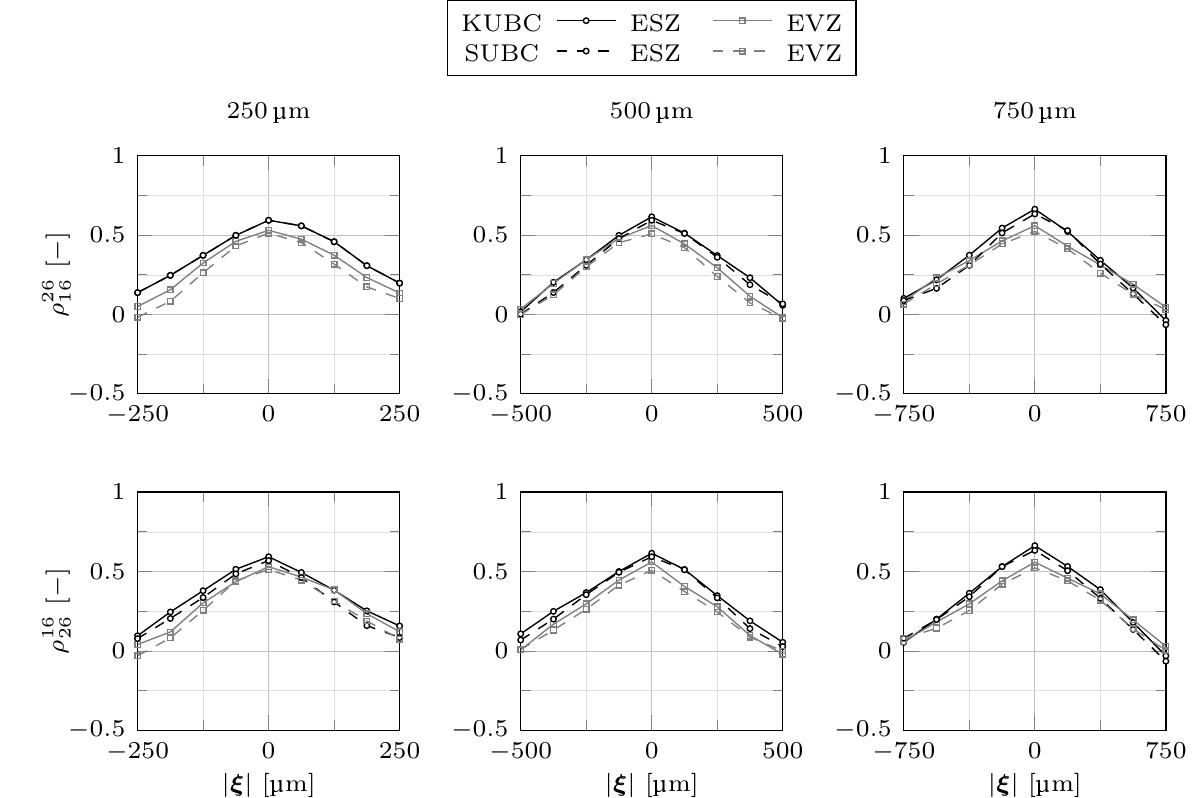}
\caption{Comparison of $\rho_{16}^{26}$ and $\rho_{26}^{16}$ assuming plane strain and plane stress states at $x = \SI{1250}{\micro\metre}$.}
\label{fig:comp3}
\end{figure}

\begin{figure}
\centering
\includegraphics[width=0.92\textwidth]{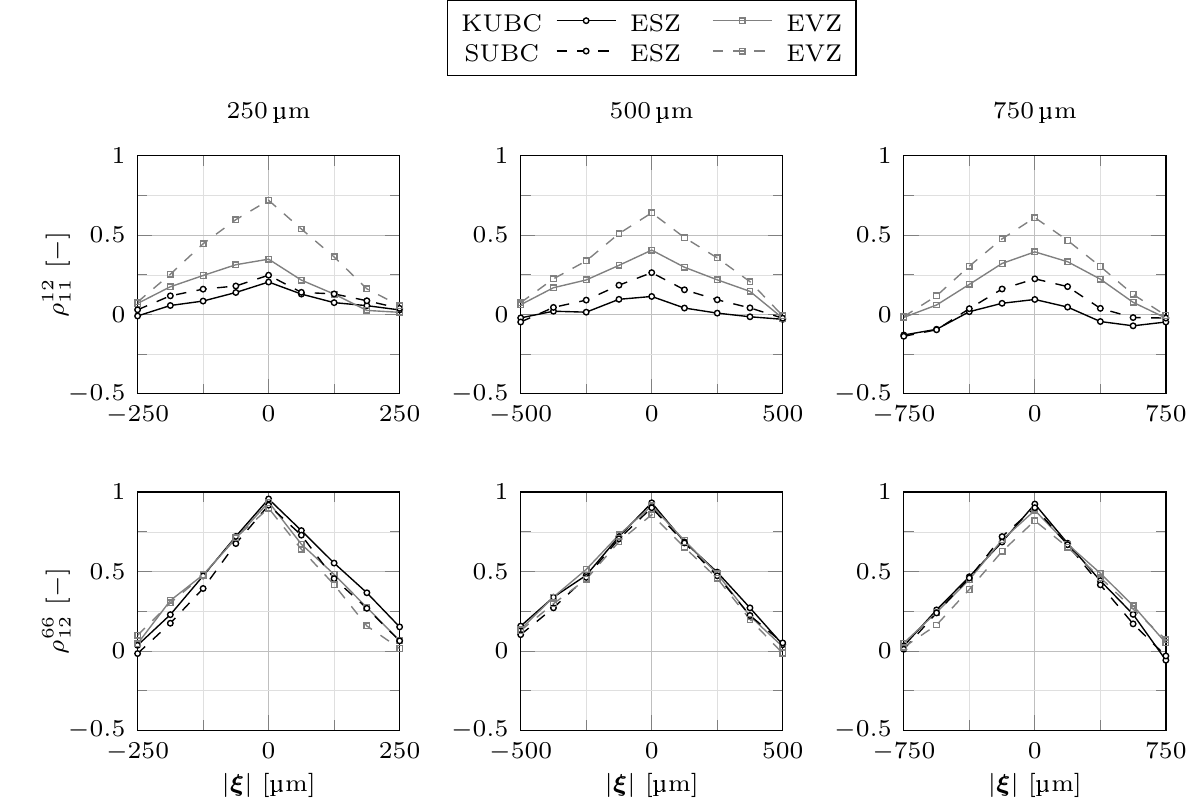}
\caption{Comparison of the dimensionless correlation parameters that are independent of the window size assuming plane strain and plane stress states at $x = \SI{1250}{\micro\metre}$.}
\label{fig:comp2}
\end{figure}

\begin{figure}
\centering
\includegraphics[width=0.92\textwidth]{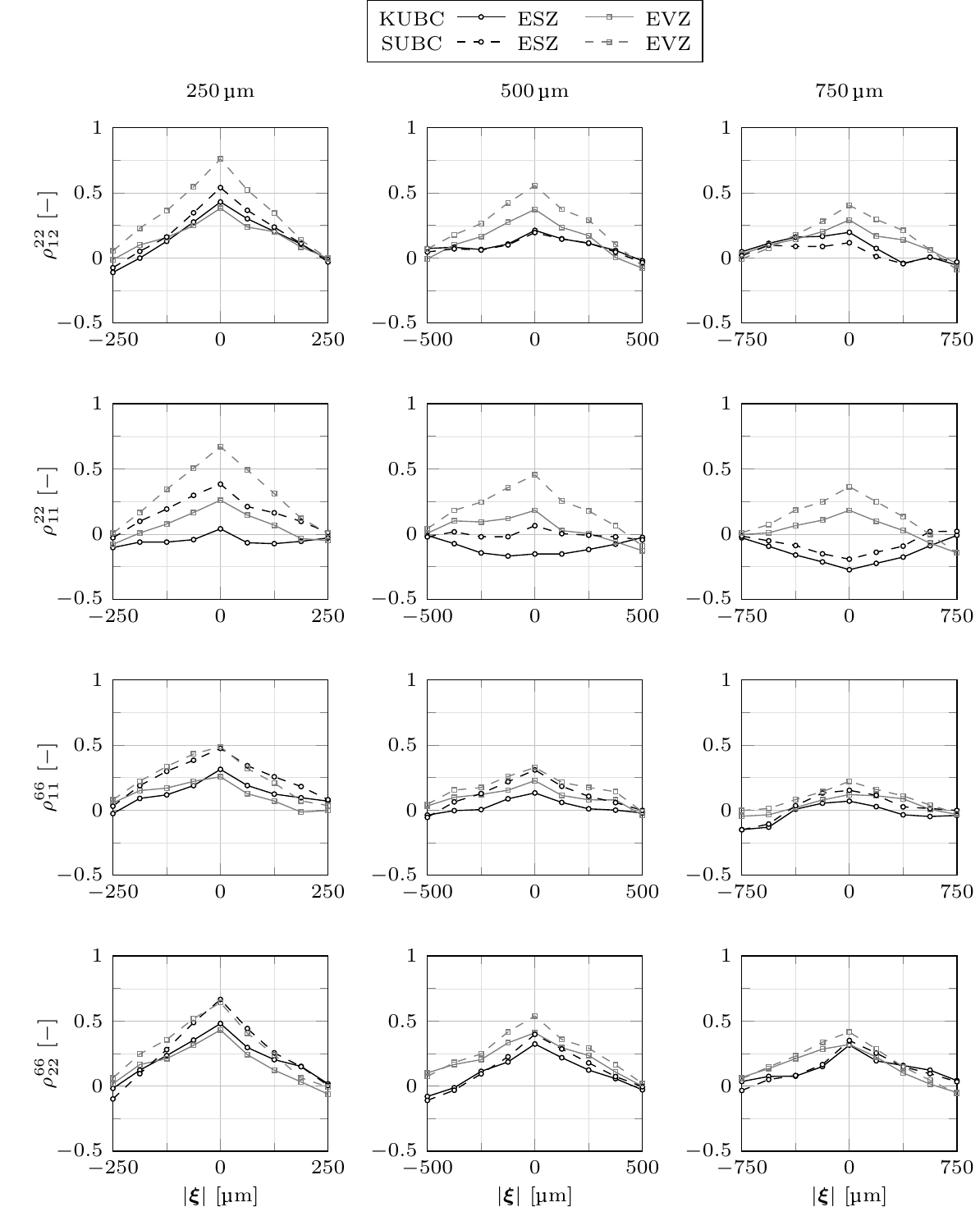}
\caption{Comparison of the dimensionless correlation parameters that depend on the window size assuming plane strain and plane stress states at $x = \SI{1250}{\micro\metre}$.}
\label{fig:comp4}
\end{figure}

\end{document}